\documentclass[12pt]{article}
\usepackage{amsfonts}

\usepackage{graphicx}
\usepackage{amsmath}
\usepackage{float}

\numberwithin{equation}{section} \numberwithin{figure}{section}

\newcommand{\ka}{\kappa}

\newcommand{\te}{\theta}

\begin{document}

\bigskip
\begin{titlepage}
\bigskip \begin{flushright}
hep-th/0210280\\
WATPPHYS-TH02/10
\end{flushright}

\vspace{1cm}

\begin{center}
{\Large \bf {Thermodynamics of $(d+1)$-dimensional NUT-charged AdS Spacetimes}}\\
\end{center}
\vspace{2cm}
\begin{center}
R. Clarkson$^{*}$\footnote{EMail: rick@avatar.uwaterloo.ca} L.
Fatibene$^{\dagger}${ \footnote{EMail: fatibene@dm.unito.it}} and
R. B. Mann$^\ddagger$ \footnote{
EMail: mann@avatar.uwaterloo.ca}\\
$^{*,\ddagger}$Department of Physics, University of Waterloo, \\
Waterloo, Ontario N2L 3G1, Canada\\
$^{\dagger}$Dipartimento di Matematica, Universit\`a di Torino, \\
via C. Alberto 10, 10123 Torino, Italy \\
\vspace{1cm}
\today\\
\end{center}

\begin{abstract}
We consider the thermodynamic properties of $(d+1)$-dimensional
spacetimes with NUT charges. Such spacetimes are asymptotically
locally anti de Sitter (or flat), with non-trivial topology in
their spatial sections, and can have fixed point sets of the
Euclidean time symmetry that are either $(d-1)$-dimensional
(called "bolts") or of lower dimensionality (pure "NUTs").  We
compute the free energy, conserved mass, and entropy for 4, 6, 8
and 10 dimensions for each, using both Noether charge methods and
the AdS/CFT-inspired counterterm approach.  We then generalize
these results to arbitrary dimensionality.  We find in $4k+2$
dimensions that there are no regions in parameter space in the
pure NUT case for which the entropy and specific heat are both
positive, and so all such spacetimes are thermodynamically
unstable.  For the pure NUT case in $4k$ dimensions a region of
stability exists in parameter space that decreases in size with
increasing dimensionality.  All bolt cases have some region of
parameter space for which thermodynamic stability can be realized.
\end{abstract}
\end{titlepage}\onecolumn

\begin{center}
\end{center}

\section{Introduction}

Ever since Beckenstein's suggestion \cite{beck} that the area of the event
horizon of a black hole is proportional to its physical entropy, the
relationship between a configuration of the gravitational field and its
thermodynamic disorder has provided fertile ground for research. \ When
combined with Hawking's demonstration \cite{hawkthermo} that black holes can
radiate with a blackbody spectrum at non-zero temperature (once quantum
effects are taken into account), the correspondence between black hole
physics and equilibrium thermodynamics becomes particularly robust. \
Despite this, the origin of black hole entropy in terms of an underlying
microphysical explanation remains inconclusive, though a number of
intriguing suggestions from both string theory and loop quantum gravity have
appeared in recent years. \

Gravitational entropy can be regarded as arising from the Gibbs-Duhem
relation applied to the path-integral formulation of quantum gravity. \
While such a formulation is undoubtedly not the last word on this subject,
at least in the semiclassical limit it should yield a relationship between
gravitational entropy and other relevant thermodynamic quantities. \ In
employing this relation, Hawking and Hunter \cite{hawkhunt} proposed that
gravitational entropy arises whenever it is not possible to foliate a given
spacetime in the Euclidean regime by a family of surfaces of constant time.
This situation can occur in $\left( d+1\right) $-dimensions if the topology
of the Euclidean spacetime is not trivial -- specifically when the
(Euclidean) timelike Killing vector $\xi =\partial /\partial \tau $\ that
generates the U(1) isometry group has a fixed point set of even
co-dimension. If this co-dimension is $\left( d-1\right) $ then the usual
relationship between area and entropy holds. However if the co-dimension is
smaller than this the relationship between area and entropy is generalized.

Such situations occur in spacetimes containing NUT-charges. \ In four
dimensions they not only can have $2$-dimensional fixed point sets (called
``bolts''), they also have $0$-dimensional fixed point sets (called
``nuts''). \ Here the orbits of the $U(1)$ isometry group develop
singularities, of dimension 1 in the orbit space, and of dimension $2$ in
the Euclidean spacetime. These singularities are the gravitational analogues
of Dirac string singularities and are referred to as Misner strings. \ When
the NUT charge is nonzero, the entropy of a given spacetime includes not
only the entropies of the $2$-dimensional bolts, but also those of the
Misner strings.

In $4$ dimensions the resulting quantities are finite although the different
contributions typically diverge with the size of the system, due to the fact
that the solutions are not asymptotically flat but asymptotically locally
anti de Sitter. These divergences have been addressed within the framework
for conserved quantities and entropy due to Brown-York (see \cite{BrownYork}%
) by computing the quantities relative to a background spacetime of
appropriate topology; for example a bolt-spacetime relative to a
nut-spacetime \cite{hawkhunt,garf,Taub}.

The $4$-dimensional case has also been addressed in terms of N\"{o}ther's
theorem (see refs. \cite{Lagrange, Remarks} and references quoted therein).
Provided that a background is introduced, conserved quantities are
considered as quantities relative to the background, and the covariant first
order Lagrangian is used (see \cite{Taub}). In this framework the entropy is
defined \textit{\`{a} la} Clausius as the quantity that satisfies the first
law of thermodynamics. In later work the relationship between N\"{o}ther
currents of the covariant first order Lagrangian and the Brown-York
framework was analyzed (see \cite{OurBrownYork}).

However, it has also been shown that such quantities can be computed
intrinsically, without reference to a background, for both non-rotating \cite%
{MannMisner} and rotating \cite{nutkerr} cases. This can be done for
asymptotically anti de Sitter (AdS) NUT-charged spacetimes by employing the
AdS/CFT-inspired counterterm approach \cite{bala}, \ and for their
asymptotically flat counterparts via a suitable generalization of this
scheme \cite{MannMisner}. \ The resultant thermodynamic behaviour has some
surprising features. \ Although the entropy can be intrinsically computed it
is no longer positive definite for all possible values of the NUT charge.
Instead there exists a finite range of values for the NUT charge for which
the entropy and specific heat are both positive; outside of this range at
least one of these quantities is negative. This occurs for both non-rotating %
\cite{MannMisner,EJM} and rotating \thinspace \cite{nutkerr} solutions.

Motivated by these unusual results, we consider in this paper an analysis of
the thermodynamic behaviour of higher--dimensional NUT -- charged
spacetimes. \ These spacetimes were first considered by Bais and Batenberg %
\cite{BaisBat} in the asymptotically flat case, and then generalized to
asymptotically AdS situations by Awad and Chamblin \cite{AC}. Despite the
fact that the two methods based on the N\"{o}ther theorem and the
counterterm approach share \textit{a priori} very few features, we show they
provide the same results. \ An off-shell comparison between these two
approaches will be postponed to a future paper.

We compute the action, conserved mass and thermodynamic quantities for
Taub-NUT-AdS and Taub-Bolt-AdS spacetimes in six, eight and ten dimensions,
and consider the general properties of the $(d+1)$-dimensional case. We find
the qualitative thermodynamic properties of NUT-charged spacetimes alternate
between dimensionalities of $4k$ and $4k+2$. For example all of the
six-dimensional quantities have an overall negative sign to their four
dimensional counterparts, whereas those in eight dimensions have the same
overall sign. The bolt cases do not share this property. We then extend our
study to the general $(d+1)$-dimensional case, and find general expressions
for the action, the conserved mass, the entropy and the specific heat for
the NUT and bolt cases. \

\section{Conserved Charges}

For later convenience we briefly summarize the N\"{o}ther and counterterm
approaches (see \cite{Taub,Remarks} and \cite{MannMisner} for details).

The N\"{o}ther framework is based on the covariant first order Lagrangian
\begin{equation}
L=\hbox{$1\over 2\ka$}[(R-\Lambda )\sqrt{g}+\hbox{d}_{\mu }(w_{\alpha \beta
}^{\mu }g^{\alpha \beta }\sqrt{g})-(\bar{R}-\Lambda )\sqrt{\bar{g}}]\>%
\hbox{d}s  \label{FOL}
\end{equation}%
where $g_{\alpha \beta }$ is the dynamical metric, $\bar{g}_{\alpha \beta }$
is the background, $R$ and $\bar{R}$ are the scalar curvatures of the metric
and the background respectively, $\Lambda =-{\frac{d(d-1)}{2\ell ^{2}}}$ is
the cosmological constant, and $w_{\alpha \beta }^{\mu }=u_{\alpha \beta
}^{\mu }-\bar{u}_{\alpha \beta }^{\mu }$ is a tensorial quantity defined by
\begin{equation}
u_{\alpha \beta }^{\mu }=\Gamma _{\alpha \beta }^{\mu }-\delta _{(\alpha
}^{\mu }\Gamma _{\beta )\lambda }^{\lambda }\qquad \bar{u}_{\alpha \beta
}^{\mu }=\bar{\Gamma}_{\alpha \beta }^{\mu }-\delta _{(\alpha }^{\mu }\bar{%
\Gamma}_{\beta )\lambda }^{\lambda }
\end{equation}%
The constant $\kappa$ depends on unit conventions and it becomes relevant
when coupling to matter fields. In this paper we shall set $\kappa=4\pi$.

We remark that the Lagrangian (\ref{FOL}) is the sum of three terms
\begin{equation}
L_1=\hbox{$1\over 2\ka$}(R-\Lambda )\sqrt{g}\>\hbox{d}s \quad L_2=%
\hbox{$1\over 2\ka$}\hbox{d}_{\mu }(w_{\alpha \beta }^{\mu }g^{\alpha \beta }%
\sqrt{g})\>\hbox{d}s \quad L_3=-\hbox{$1\over 2\ka$}(\bar{R}-\Lambda )\sqrt{%
\bar{g}}\>\hbox{d}s  \label{FOLa}
\end{equation}
each of which is a covariant Lagrangian on its own. Hence in a geometrical
well-defined way, the total conserved quantities $Q$ associated with the
total Lagrangian $L$ will split as $Q= Q_1+Q_2+Q_3$, each term being
associated to a partial Lagrangian $L_1$, $L_2$, $L_3$, respectively.

We also remark that $\hbox{d}s$ is the standard local volume element
associated with the coordinates used on the spacetime $M$ of dimension $d+1$%
. The equality $\hbox{d}s= \hbox{d}^{d+1}x$ holds. The double notation is
maintained for coherence with the quoted references.

The N\"{o}ther theorem, when applied to this Lagrangian, has proved to be
effective in solving the anomalous factor problems \cite{Katz} known for the
Komar potential (see \cite{Komar}). By choosing a suitable background it has
been shown to be effective in the case of non-asymptotically flat solutions
generalizing the standard ADM conserved quantities.

The conserved quantities are defined as the integral of the following
superpotential
\begin{equation}
\mathcal{U}(\xi )=\hbox{$1\over 2\ka$}[\nabla ^{\alpha }\xi ^{\beta }\sqrt{g}%
+\xi ^{\alpha }w_{\mu \nu }^{\beta }g^{\mu \nu }\sqrt{g}-\bar{\nabla}%
^{\alpha }\xi ^{\beta }\sqrt{\bar{g}}]\>\hbox{d}s_{\alpha \beta }
\label{superpotential}
\end{equation}%
for any spacetime vector field $\xi $. We remark that this superpotential
inherits a splitting $\mathcal{U}(\xi )=\mathcal{U}_{1}(\xi )+\mathcal{U}%
_{2}(\xi )+\mathcal{U}_{3}(\xi )$ from the splitting of the Lagrangian, i.e.
\begin{eqnarray}
L_{1} &\rightarrow &\mathcal{U}_{1}(\xi )=\hbox{$1\over
2\ka$}\nabla ^{\alpha }\xi ^{\beta }\sqrt{g}\>\hbox{d}s_{\alpha \beta }
\notag \\
L_{2} &\rightarrow &\mathcal{U}_{2}(\xi )=\hbox{$1\over
2\ka$}\xi ^{\alpha }w_{\mu \nu }^{\beta }g^{\mu \nu }\sqrt{g}\>\hbox{d}%
s_{\alpha \beta } \\
L_{3} &\rightarrow &\mathcal{U}_{3}(\xi )=-\hbox{$1\over 2\ka$}\bar{\nabla}%
^{\alpha }\xi ^{\beta }\sqrt{\bar{g}}\>\hbox{d}s_{\alpha \beta }  \notag
\end{eqnarray}%
By specifying $\xi $ the energy-momentum and the angular momentum of the
spacetime with metric $\ g$ relative to the background metric $\bar{g}$ are
obtained. Here $\nabla ^{\alpha }$ and $\bar{\nabla}^{\alpha }$ are the
covariant derivatives with respect to the dynamical metric and the
background respectively.

The superpotential $\mathcal{U}(\xi)$ (computed along the dynamical and the
background metric) defines the conserved current (relative to the background
$\bar g$) within a closed $(d-1)$-dimensional submanifold $D$ of spacetime
\begin{equation}
Q_D(\xi)= \int_D \mathcal{U}(\xi, g, \bar g)
\end{equation}

In the limit that $D$ bounds a leaf of the foliation of the full spacetime
(e.g. $r\rightarrow \infty $ in Minkowski) one obtains the global conserved
quantities $Q_{D_{\infty }}(\xi )$. For a fixed ADM slicing a time
translation generates the mass $M$ while an asymptotic rotation generates
the angular momentum. By imposing suitable boundary conditions one can also
define the conserved quantity in a finite spacelike region \cite%
{OurBrownYork}.

The entropy is defined when the relevant thermodynamic potentials are
provided. In particular the temperature $T=\beta ^{-1}$ has to be provided
by some other physical means (e.g. from the radiation spectrum of a black
hole). In general the momenta conjugate to the angular momentum and the
other (gauge) charges have to be provided as well. For Taub-bolt (NUT)
solutions the (inverse) temperature is identified with the period of the
Euclidean time, chosen so that the Euclidean manifold is regular at all
degeneracies of the foliation.

For a family of solutions of the Taub-bolt type described below, the first
law of thermodynamics is of the following form
\begin{equation}
\delta M=T\delta S
\end{equation}
from which we can obtain the entropy $S$ by integration. This method,
introduced in \cite{Wald} and generalized in \cite{Remarks, Taub}, proved to
be effective in many situations (see also \cite{BCM,BTZ,DasMann}).

Notice that since the entropy is obtained as an integral quantity, in this
context it is defined modulo a constant not depending on the solution under
investigation but possibly depending on the model (e.g. on the cosmological
constant).

Turning next to the counterterm approach we modify the above action in $d+1$
dimensions to be
\begin{equation}
I=I_{B}+I_{\partial B}+I_{ct}  \label{action}
\end{equation}%
where
\begin{eqnarray}
I_{B} &=&\frac{1}{16\pi G}\int_{\mathcal{M}}d^{d+1}x\sqrt{-g}\left(
R-2\Lambda +\mathcal{L}_{M}\left( \Psi \right) \right)   \label{actbulk} \\
I_{\partial B} &=&-\frac{1}{8\pi G}\int_{\mathcal{\partial M}}d^{d}x\sqrt{%
-\gamma }\Theta   \label{actbound}
\end{eqnarray}%
and $\mathcal{L}_{M}$ is the Lagrangian for matter fields $\Psi $. \ The
first term in (\ref{action}) is the bulk action over the $d+1$ dimensional
Manifold $\mathcal{M}$ with metric $g$ and the second term (\ref{actbound})
is a surface term necessary to ensure that the Euler-Lagrange variation is
well-defined, yielding the Einstein equations of motion with a positive
cosmological constant. The boundary $\mathcal{\partial M}$ (with induced
metric $\gamma $) of the manifold in general consists of both spacelike and
timelike hypersurfaces. \ For an asymptotically AdS spacetime it will be the
Einstein cylinder at infinity; for an asymptotically dS spacetime it will be
the union of spatial Euclidean boundaries at early and late times.

The term $I_{ct}$ is due to the contributions of the counterterms from the
boundary CFT \cite{bala}. The existence of these terms is suggested by the
AdS/CFT correspondence conjecture, which posits the following relationship
\begin{equation}
Z_{AdS}[\gamma ,\Psi _{0}]=\int_{[\gamma ,\Psi _{0}]}D\left[ g\right] D\left[
\Psi \right] e^{-I\left[ g,\Psi \right] }=\left\langle \exp \left(
\int_{\partial \mathcal{M}{_{d}}}d^{d}x\sqrt{g}\mathcal{O}_{[\gamma ,\Psi
_{0}]}\right) \right\rangle =Z_{CFT}[\gamma ,\Psi _{0}]  \label{PAR}
\end{equation}%
between the partition function of any field theory on $AdS_{d+1}$ and a
quantum conformal field theory defined on the boundary of $AdS_{d+1}$ . The
induced boundary metric and matter fields are respectively denoted by $%
\gamma $ and $\Psi _{0}$ symbolically with $\mathcal{O}$ a quasi-primary
conformal operator defined on the boundary of $AdS_{d+1}$. The integration
is over configurations $\left[ g,\Psi \right] $ of metric and matter fields
that approach $[\gamma ,\Psi _{0}]$ when one goes from the bulk of $%
AdS_{d+1} $ to its boundary. This conjecture has been verified for several
important examples, encouraging the expectation that an understanding of
quantum gravity in a given spacetime (at least an asymptotically AdS one)
can be obtained by studying its holographic CFT dual, defined on the
boundary of spacetime at infinity.

The boundary counterterm action $I_{ct}$ arises from the counterterms of the
CFT. It is universal, depending only on curvature invariants that are
functionals of the intrinsic boundary geometry, leaving the equations of
motion unchanged. By rewriting the Einstein equations in Gauss-Codacci form,
and then solving them in terms of the extrinsic curvature functional of the
boundary $\mathcal{\partial M}$ and its normal derivatives to obtain the
divergent parts \cite{KLS}, it can be generated by an algorithmic procedure.
The entire divergent structure can be covariantly isolated for any given
boundary dimension $d$, since all divergent parts can be expressed in terms
of intrinsic boundary data and do not depend on normal derivatives \cite%
{Feffgraham}. By varying the boundary metric under a Weyl transformation, it
can be shown that the trace $\tilde{\Pi}$ of the extrinsic curvature is
proportional to the divergent boundary counterterm Lagrangian. No background
spacetime is required, and computations of the action and conserved charges
yield unambiguous finite values intrinsic to the spacetime, as has been
verified in numerous examples \cite{MannMisner,nutkerr,EJM}.

The result of this procedure is that
\begin{eqnarray}
\mathcal{L}_{\text{ct}} &=&\int d^{d}x\sqrt{-\gamma }\left\{ -\frac{d-1}{
\ell }-\frac{\ell \mathsf{\Theta }\left( d-3\right) }{2(d-2)}\mathsf{R}-
\frac{\ell ^{3}\mathsf{\Theta }\left( d-5\right) }{2(d-2)^{2}(d-4)}\left(
\mathsf{R}_{ab}\mathsf{R}^{ab}-\frac{d}{4(d-1)}\mathsf{R}^{2}\right) \right.
\notag \\
&&+\frac{\ell ^{5}\mathsf{\Theta }\left( d-7\right) }{(d-2)^{3}(d-4)(d-6)}
\left( \frac{3d+2}{4(d-1)}\mathsf{RR}^{ab}\mathsf{R}_{ab}-\frac{d(d+2)}{
16(d-1)^{2}}\mathsf{R}^{3}\right.  \notag \\
&&\left. -2\mathsf{R}^{ab}\mathsf{R}^{cd}\mathsf{R}_{acbd}+\left. -\frac{d}{
4(d-1)}\nabla _{a}\mathsf{R}\nabla ^{a}\mathsf{R}+\nabla ^{c}\mathsf{R}
^{ab}\nabla _{c}\mathsf{R}_{ab}\right) \right\}  \label{Lagrangianct}
\end{eqnarray}
where $\Lambda =-3/\ell ^{2}$\ and $\mathsf{R}$ is the curvature associated
with the induced metric $\gamma $. \ The series truncates for any fixed
dimension, with new terms entering at every new even value of $d$, as
denoted by the step-function ($\mathsf{\Theta }\left( x\right) =1$ provided $%
x\geq 0$, and vanishes otherwise).\ As is the case for its AdS counterpart,
conserved charges on the spatially infinite boundaries of an asymptotically
dS spacetime can be defined using (\ref{Mcons}).

Taking the variation of the action (\ref{action}) and carefully keeping
account of all boundary terms, a conserved charge
\begin{equation}
\mathfrak{Q}_{\xi }=\oint_{\Sigma }d^{d-1}S^{a}~\xi ^{b}T_{ab}^{{eff}}
\label{Mcons}
\end{equation}%
can be associated with a closed surface $\Sigma $ (with normal $n^{a}$),
provided the boundary geometry has an isometry generated by a Killing vector
$\xi ^{\mu }$. The quantity $T_{ab}^{{eff}}$ is given by the variation of
the action (\ref{action}) at the boundary with respect to $\gamma ^{ab}$,
and the quantity $\mathfrak{Q}_{\xi }$ is conserved between closed surfaces $
\Sigma $\ distinguished by some foliation parameter $\tau $. \ If $\xi
=\partial /\partial t$ then $\mathfrak{Q}$ is the conserved mass/energy $%
\mathfrak{M}$; if $\xi _{a}=\partial /\partial \phi ^{a}$ then $\mathfrak{Q}$
is the conserved angular momentum ${J}$ provided $\phi $ is a periodic
coordinate associated with $\Sigma $. \ Details of this formulation can be
found in refs. \cite{BrownYork,BCM,ivan}.

We can proceed further by formulating gravitational thermodynamics via the
Euclidean path integral
\begin{equation}
Z=\int D\left[ g\right] D\left[ \Psi \right] e^{-I\left[ g,\Psi \right]
}\simeq e^{-I_{{cl}}}  \label{Zpath}
\end{equation}%
where one integrates over all metrics and matter fields between some given
initial and final Euclidean hypersurfaces, taking $\tau $ to have some
period $\beta _{H}$, determined by demanding regularity of the Euclideanized
manifold at degenerate points of the foliation. Semiclassically the result
is given by the classical action evaluated on the equations of motion, and
yields to this order
\begin{equation}
S=\beta _{H}\mathfrak{M}_{\infty }-I_{{cl}}  \label{entropy}
\end{equation}%
upon application of the Gibbs-Duhem relation to the partition function. \
The gravitational entropy $S$ is defined to be the difference between the
total energy at infinity $\mathfrak{M}_{\infty }$ and the free energy $%
\mathfrak{F}=I_{{cl}}/\beta _{H}$ multiplied by $\beta _{H}$, which can be
interpreted as the inverse temperature. \ This quantity will be non-zero
whenever there is a mismatch between $\mathfrak{M}$ and $\mathfrak{F}$. \

Note that in this approach there is no freedom to shift the value of the
entropy by a constant. We shall find in the sequel that all Bolt
AdS-spacetimes have an additional constant independent of the NUT charge.
This constant has no effect on the first law of thermodynamics (in which
only changes of entropy are observable); its physical value is such that the
entropy vanishes for the ground state (vanishing NUT charge).

In the sequel we shall analyze the conservation laws of Taub-NUT/bolt
solutions in dimension $6$, $8$ and $10$, both by N\"{o}ther theorem and
counterterm methods. We stress that there is no \textit{a priori} reason for
these methods to provide the same results, though, as we shall see, in this
case the methods completely agree. The entropy is computed, both by the
Gibbs-Duhem relation and by the first law of thermodynamics. These two
methods provide two expressions for the entropy which are in agreement apart
from the aforementioned integration constant.

\bigskip

\section{Four Dimensional Taub-NUT Spacetimes}

We begin by briefly recapitulating the results from four dimensions \cite%
{MannMisner}, \cite{EJM}, and \cite{CEJM}. The metric is given by
\begin{equation}
ds^{2}=F(r)\left( d\tau +2N\cos (\theta )d\phi \right)^2 +\frac{dr^{2}}{F(r)}
+(r^{2}-N^{2})\left( d\theta ^{2}+\sin ^{2}(\theta )d\phi ^{2}\right)
\label{4dmetric}
\end{equation}
with
\begin{equation}
F(r)=\frac{r^{2}+N^{2}-2mr+\ell ^{-2}(r^{4}-6N^{2}r^{2}-3N^{4})}{r^{2}-N^{2}}
\label{Fr4d}
\end{equation}
The general action was found to be
\begin{equation}
I_{4}=\frac{\beta }{2\ell ^{2}}\left( \ell ^{2}m+3N^{2}r_{+}-r_{+}^{3}\right)
\label{4ditot}
\end{equation}
where $\beta $ is the period of $\tau $, given in four dimensions by
\begin{equation}
\beta =8\pi N  \label{bet4}
\end{equation}%
and determined by demanding regularity of the manifold so that the
singularities at $\theta =0,\pi $ are coordinate artifacts.

There is an additional regularity criterion to be satisfied, namely the
absence of conical singularities at the roots of the function $F(r)$. This
can be ensured by demanding that the period of of $\tau $ is
\begin{equation}
\beta =\left| \frac{4\pi }{F^{\prime }\left( r_{+}\right) }\right| =8\pi N
\label{bet4a}
\end{equation}
where $F(r=r_{+})=0$, and the second equality follows by demanding
consistency with eq. (\ref{bet4}). \ It is straightforward to show that
there are only two solutions to eq. (\ref{bet4a}), one where $r_{+}=N$ and
one where $r_{+}=r_{b}>N$, referred to respectively as the NUT and Bolt
solutions. In the former case we have
\begin{equation}
m_{n}=\frac{N(\ell ^{2}-4N^{2})}{\ell ^{2}}  \label{nut4mass}
\end{equation}%
whereas solving (\ref{bet4a}) in this latter case gives two possible
solutions
\begin{equation}
r_{b\pm }=\frac{\ell ^{2}\pm \sqrt{\ell ^{4}-48N^{2}\ell ^{2}+144N^{4}}}{12N}
\label{rbolt4}
\end{equation}
with the limit on $N$ being
\begin{equation}
N\leq \frac{(3\sqrt{2}-\sqrt{6})\ell }{12}=N_{max}  \label{Nlimit4}
\end{equation}
since we want the discriminant to be real.

The actions can be calculated for these two solutions, using the NUT and
Bolt masses (found by solving $F(r=N)=0$ for $m=m_{n}$ and $F(r=r_{b})=0$
for $m=m_{b}$; see tables \ref{summarytableNUT}, \ref{summarytableBolt} at
the end of the paper).

The actions in four dimensions are
\begin{eqnarray}
I_{NUT} &=&\frac{4\pi N^{2}(\ell ^{2}-2N^{2})}{\ell ^{2}}  \label{INUT4d} \\
I_{Bolt} &=&\frac{-\pi (r_{b}^{4}-\ell ^{2}r_{b}^{2}+N^{2}(3N^{2}-\ell
^{2})) }{3r_{b}^{2}-3N^{2}+\ell ^{2}}  \label{IBolt4d}
\end{eqnarray}
The entropy and specific heat for the NUT and Bolt solutions have also been
calculated \cite{CEJM}, \cite{MannMisner}:
\begin{eqnarray}
S_{NUT} &=&\frac{4\pi N^{2}(\ell^2 -6N^{2})}{\ell ^{2}} \\
C_{NUT} &=&\frac{8\pi N^{2}(-\ell^{2}+12N^{2})}{\ell^{2}}
\end{eqnarray}
Note that that the entropy becomes negative for $N>{\frac{\ell }{\sqrt{6}}}$ %
\cite{EJM,MannMisner} and the specific heat negative for $N<{\frac{\ell }{%
\sqrt{12}}}$. Thus, for thermally stable solutions, the value of the NUT
charge must be in the range
\begin{equation}
\frac{\ell }{\sqrt{12}}\leq N\leq \frac{\ell }{\sqrt{6}}
\end{equation}
For the Bolt, the entropy and specific heat are found to be
\begin{eqnarray}
S_{Bolt} &=&\frac{\pi
(3r_{b}^{4}+(\ell^{2}-12N^{2})r_{b}^{2}+N^{2}(\ell^{2}-3N^{2}))}{%
3r_{b}^{2}+\ell^{2}-3N^{2}} \\
C_{Bolt}(r_{b}=r_{b\pm }) &=&\frac{\pi \ell^{4}}{36N^{2}}\pm \frac{\pi
(\ell^{8}-24N^{2}\ell^{6}+144\ell^{4}N^{4}-10368N^{6}\ell^{2}+41472N^{8}) }{%
36N^{2}\ell^{2}\sqrt{\ell^{4}+144N^{4}-48N^{2}\ell^{2}}}
\end{eqnarray}
where we recall in $S_{Bolt}\,$that $r_{b}=r_{b\pm }$; $C_{Bolt}$ has been
given explicitly for each $r_{b\pm }$. A plot of the entropy and specific
heat vs. $N/\ell $ indicates that the upper branch solutions ($r_{b}=r_{b+}$%
) are thermally stable, whereas the lower branch solutions ($r_{b}=r_{b-}$)
are thermally unstable. As these plots are qualitatively similar to the
six-dimensional case (see figures \ref{6dSCBoltplot}, \ref{6dSBmPlot}, \ref%
{6dCBmPlot}), we shall not reproduce them here.

The analysis via N\"other theorem was carried out assuming no cosmological
constant (see \cite{Taub}). Here we shall generalize to the case of non-zero
cosmological case. We set $\xi=\partial_t+ a\partial_\te$ and we compute the
relative energy between Taub-bolt and Taub-NUT solution within the spatial
sphere $\{\tau=\tau_0, r=r_0\}$. We obtain $Q_\infty(\xi)= Q_1+Q_2+ Q_3$
where
\begin{eqnarray}
Q_1 & = & \frac{\pi}{\ell^2\kappa}\left( 2r_0^3 -2N^2 r_0 +\frac{r_b^4+
r_b^2(\ell^2-6N^2)+N^2(\ell^2-3N^2)}{r_b} \right) + O(r_0^{-1}) \\
Q_2 & = & \frac{\pi}{\ell^2\kappa}\left( \frac{r_b^4+ r_b^2 (\ell^2-6N^2)
+2N(4N^2-\ell^2) r_b +N^2(\ell^2-3N^2)}{r_b} \right) + O(r_0^{-1}) \\
Q_3& = & \frac{\pi}{\ell^2\kappa}\left( -2r_0^3 +2N^2 r_0 -2N( \ell^2-4N^2)
\right)+ O(r_0^{-1})
\end{eqnarray}
The total relative energy in the limit $r_0\rightarrow \infty$ is
\begin{equation}
Q_\infty(\xi)= \frac{2\pi}{\ell^2\kappa}\left( r_b^3 +(-6N^2+\ell^2)r_b
-2N(\ell^2-4N^2) +\frac{N^2(\ell^2-3N^2)}{r_b} \right)
\end{equation}
Hence we have $Q_\infty(\xi)= \frac{4\pi}{\kappa}(m_b- m_n)$ and no angular
momentum, as expected.

Assuming the temperature $T=\beta^{-1}=(8\pi N)^{-1}$ the resulting entropy
is
\begin{equation}
S= \frac{2\pi^2}{\ell^2\kappa}\left( \frac{(\ell^4-144N^4)r_b}{3N}
+4N^2(12N^2-\ell^2) \right)
\end{equation}

Hence, comparing to the results above, we obtain
\begin{equation}
S=\frac{4\pi }{\kappa }\left(S_{Bolt}-S_{NUT} +\frac{\pi\ell^2}{3}\right)
\end{equation}
Hence the same classical entropy is obtained by the two methods up to an
integration constant.

\section{Six Dimensional Solutions}

In six dimensions, there are two possible forms of the metric (as given by %
\cite{AC}). One form uses $S^{2}\times S^{2}$ as a base space, and has the
following form;
\begin{eqnarray}
ds^{2} &=&F(r)\left( d\tau +2N\cos (\theta _{1})d\phi _{1}+2N\cos (\theta
_{2})d\phi _{2}\right) ^{2}+\frac{dr^{2}}{F(r)}+(r^{2}-N^{2})(d\theta
_{1}^{2}  \notag \\
&&+\sin ^{2}(\theta _{1})d\phi _{1}^{2}+d\theta _{2}^{2}+\sin ^{2}(\theta
_{2})d\phi _{2}^{2})  \label{6dmetric1}
\end{eqnarray}
The second form uses $\mathcal{B}=\mathbb{CP}^{2}$ as a base space, and
generalizes the metric to include a negative cosmological constant. It has
the form,
\begin{equation}
ds^{2}=F(r)(d\tau +A)^{2}+\frac{dr^{2}}{F(r)}+(r^{2}-N^{2})d\Sigma _{2}^{~2}
\label{6dmetricCPForm}
\end{equation}
where $d\Sigma _{2}^{~2}$ is the metric for the $\mathbb{CP}^{2}$ space,
with the form
\begin{equation}
d\Sigma _{2}^{~2}=\frac{du^{2}}{\left( 1+\frac{u^{2}}{6}\right) ^{2}}+\frac{%
u^{2}}{4\left( 1+\frac{u^{2}}{6}\right) ^{2}}\left( d\psi +\cos (\theta
)d\phi \right) ^{2}+\frac{u^{2}}{4\left( 1+\frac{u^{2}}{6}\right) }(d\theta
^{2}+\sin ^{2}(\theta )d\phi ^{2})  \label{CP2Metric}
\end{equation}
and the one-form $A$ is given by
\begin{equation}
A=\frac{u^{2}N}{2\left( 1+\frac{u^{2}}{6}\right) }\left( d\psi +\cos (\theta
)d\phi \right)
\end{equation}

Both forms of the metric use the same form for $F(r)$, given in six
dimensions by
\begin{equation*}
F(r)=\frac{3r^{6}+(\ell ^{2}-15N^{2})r^{4}-3N^{2}(2\ell
^{2}-15N^{2})r^{2}-6mr\ell ^{2}-3N^{4}(\ell ^{2}-5N^{2})}{3\ell
^{2}(r^{2}-N^{2})^{2}}
\end{equation*}
and yield exactly the same action and and total mass, the only difference
between the calculations being that each base space has a different volume
element associated with it.

Using the method of counter-terms, before specializing to NUT or Bolt
solutions, the action is found to be
\begin{equation}
I=\frac{2\pi \beta (-3r_{+}^{5}+10N^{2}r_{+}^{3}-15N^{4}r_{+}+3m\ell^{2})}{
3\ell^{2}}  \label{6dItot}
\end{equation}
with $r_{+}$ the value of $r$ that is the largest positive root of $F\left(
r\right) $, determined by the fixed point set of $\partial _{\tau }$. The
quantity $\beta $ is the period of $\tau $, which in 6 dimensions is given by

\begin{equation}
\beta =12\pi N  \label{betagen}
\end{equation}
The flat space limit of the total action (\ref{6dItot}) can be calculated by
substituting in the largest positive root of $F(r)$; this value is given by $%
r_{b-}$ (\ref{6drbpm}), and yields

\begin{equation}
I\rightarrow 2m\pi \beta
\end{equation}

As the metric is not rotating we have only the conserved total mass/energy $%
\mathfrak{M}$, which from (\ref{Mcons}) is
\begin{equation}
\mathfrak{M}_{\xi }=\frac{1}{8\pi }\oint_{\partial M_{\infty }\bigcap \Sigma
_{\tau }}T_{\mu \nu }u^{\mu }\xi ^{\nu }  \label{consQ}
\end{equation}
where $T_{\mu \nu }$ is the boundary stress-energy (see \cite{DasMann}), $%
u^{\mu }$ is the timelike unit vector, and $\xi ^{\mu }=\left[ 1,0,0,\ldots %
\right] $ is the timelike killing vector. Inserting these into the integral,
along with $T_{\mu \nu }$ for six dimensions, $\mathfrak{M}_{\xi }$ becomes
\begin{equation}
\mathfrak{M}_{\xi }=\frac{1}{8\pi }(4\pi )^{2}\int_{0}^{\pi }d\theta
_{1}\int_{0}^{\pi }d\theta _{2}\sqrt{\gamma }\left[ \Theta _{\mu \nu
}-\gamma _{\mu \nu }\Theta -\frac{(d-1)}{\ell }\gamma _{\mu \nu }+\ldots %
\right] u^{\mu }\xi ^{\nu }  \notag
\end{equation}
which, upon explicit calculation will yield
\begin{equation}
\mathfrak{M}=8\pi m  \label{6dM}
\end{equation}
where $\Theta _{\mu \nu }=\gamma _{\mu }^{~\rho }\gamma _{\nu }^{~\sigma
}n_{\sigma ;\rho }$ is the extrinsic curvature, evaluated on the boundary %
\cite{DasMann}.

\subsection{Taub-NUT-AdS Solution}

When $r=N$, the fixed-point set of $\partial _{\tau }$ is 2-dimensional and
we have a NUT. This implies $F\left( N\right) =0$, yielding \ \cite{AC}
\begin{equation}
m_{n}=\frac{4N^{3}(6N^{2}-\ell ^{2})}{3\ell ^{2}}  \label{mn6d}
\end{equation}%
for the mass parameter. Note that, in comparing this to the four dimensional
NUT case done in \cite{EJM}, we see that the overall sign in (\ref{mn6d}) is
minus that of the $m_{n}$ found in four dimensions (eq. (\ref{nut4mass})) in
both the large-$\ell $ and large-$N$ limits.

Imposing regularity at the root $r=N$, we find that (\ref{betagen}) holds
for the NUT-AdS metric, and the action (\ref{6dItot}) becomes
\begin{equation}
I_{NUT}=\frac{32\pi ^{2}N^{4}(4N^{2}-\ell ^{2})}{\ell ^{2}}  \label{6dIN}
\end{equation}%
Using the Gibbs-Duhem relation (\ref{entropy}) and the expression $C=-\beta
\partial _{\beta }S$ \ for the specific heat, we obtain
\begin{eqnarray}
S_{NUT} &=&\frac{32\pi ^{2}N^{4}(20N^{2}-3\ell ^{2})}{\ell ^{2}}
\label{6dSN} \\
C_{NUT} &=&\frac{384\pi ^{2}N^{4}(\ell ^{2}-10N^{2})}{\ell ^{2}}
\label{6dCN}
\end{eqnarray}
It is straightforward to check that the above equations satisfy the first
law of black hole thermodynamics, $dS=\beta dH$.

\begin{figure}[tbp]
\centering
\begin{minipage}[c]{.55\textwidth}
         \centering
         \includegraphics[width=\textwidth]{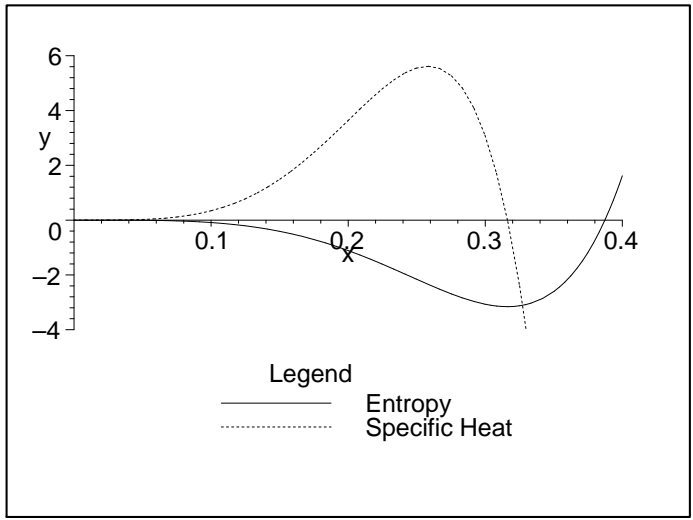}
         \caption{Plot of the NUT Entropy and Specific Heat vs. $N$ in
6 dimensions}
         \label{6dSCPlot}
     \end{minipage}
\end{figure}

We see that the action becomes negative for $N<{\frac{\ell }{2}}$. More
importantly, the entropy becomes negative for $N<\ell \sqrt{{\frac{3}{20}}}$%
, and the specific heat for $N>{\frac{\ell }{\sqrt{10}}}$. The relevance of
these values is more clearly seen in Figure \ref{6dSCPlot}, where it can be
seen that nowhere are the entropy and specific heat positive for the same
values of $N$. This means that the six dimensional Taub-NUT-AdS is
thermodynamically unstable. Of course a negative specific heat is more
palatable -- even the Schwarzschild solution has a negative specific heat. \
In situations where this is physically admissible, it means that there is an
upper bound \ $N<\ell \sqrt{{\frac{3}{20}}}$ to the NUT charge if the
entropy is to remain positive. It can also be seen in this plot that the
high temperature limits ($N\rightarrow 0$) of the entropy and specific heat
are zero; the action also goes to zero in this limit.

Finally, in the flat space limit as $\ell \rightarrow \infty $, the action,
entropy and specific heat become
\begin{eqnarray}
I_{NUT} &\rightarrow &-32\pi ^{2}N^{4}  \notag \\
S_{NUT} &\rightarrow &-96\pi ^{2}N^{4}  \label{NUT6flat} \\
C_{NUT} &\rightarrow &384\pi ^{2}N^{4}  \notag
\end{eqnarray}%
again indicating that asymptotically locally flat spacetimes with NUT charge
in six dimensions are thermodynamically unstable for all values of $N$. \

\subsection{Taub-Bolt-AdS Solution}

In the case of the Bolt solution, we require that the fixed points of $%
\partial _{\tau }$ be $4$-dimensional, implying $r_{+}=r_{b}>N$ and
\begin{equation}
m_{b}=\frac{3r_{b}^{6}+(\ell ^{2}-15N^{2})r_{b}^{4}-3N^{2}(2\ell
^{2}-15N^{2})r_{b}^{2}-3N^{4}(\ell ^{2}-5N^{2})}{6r_{b}\ell ^{2}}
\label{6dmb}
\end{equation}
The conditions \cite{AC} for a regular Bolt solution are (i) $F(r_{b})=0$
and (ii) $F^{\prime }(r_{b})={\frac{1}{3N}}$. From (ii), $r_{b}$ is
\begin{equation}
r_{b\pm }=\frac{\ell ^{2}\pm \sqrt{\ell ^{4}-180N^{2}\ell ^{2}+900N^{4}}}{%
30N }  \label{6drbpm}
\end{equation}
where reality requirements imply
\begin{equation}
N\leq \left( \frac{\sqrt{15}}{15}-\frac{\sqrt{30}}{30}\right) \ell =N_{\text{
max}}  \label{N6dlimit}
\end{equation}
since $r_{b}>N$.

We obtain from (\ref{6dItot})
\begin{equation}
I_{Bolt}=\frac{-4\pi ^{2}(3r_{b}^{6}-(5N^{2}+\ell
^{2})r_{b}^{4}-3N^{2}(5N^{2}-2\ell ^{2})r_{b}^{2}+3N^{4}(\ell ^{2}-5N^{2}))}{
3(5r_{b}^{2}+\ell ^{2}-5N^{2})}  \label{6dIB}
\end{equation}
for the Bolt action, where regularity requires
\begin{equation}
\beta _{Bolt}=\frac{4\pi }{F^{\prime }(r_{b})}=\frac{6\pi \ell
^{2}(r_{b}^{2}-N^{2})^{3}}{3r_{b}^{7}-9r_{b}^{5}N^{2}+N^{2}(4\ell
^{2}-15N^{2})r_{b}^{3}+9m\ell ^{2}r_{b}^{2}+3N^{4}(4\ell
^{2}-25N^{2})r_{b}+3m\ell ^{2}N^{2}}  \label{6dbetaBolt}
\end{equation}
and the temperature for the two solutions is the same. In fact by
substituting the expression of $r_{b}$ and $m_{b}$ into $\beta _{Bolt}$ one
readily obtains $\beta _{Bolt}=12\pi N$.

The Bolt entropy is then
\begin{equation}
S_{Bolt}=\frac{4\pi ^{2}(15r_{b}^{6} - (65N^{2}-3\ell^{2})r_{b}^{4} +
3N^{2}(55N^{2}- 6\ell^{2})r_{b}^{2} + 9N^{4}(5N^{2} - \ell^{2})) }{
3(5r_{b}^{2} + \ell^{2} - 5N^{2})}  \label{6dSB}
\end{equation}
and it is straightforward (albeit tedious) to check that the first law is
satisfied. However since $r_{b}=r_{b\pm }(N)$, there are two ``branches''
for each quantity, and these must each be separately checked.

In computing the specific heat for the Bolt we obtain a longer expression.
The two branches for the specific heat are:
\begin{eqnarray}
C_{Bolt}(r_{b}=r_{b\pm }) &=&\frac{-8\pi ^{2}}{50625}\Bigg[\frac{
\ell^{6}(90N^{2}-\ell^{2})}{N^{4}}  \label{6dCB} \\
&&\pm ~~\Bigg((\ell^{2}+30N^{2})(-\ell^{2}+30N^{2})\times  \notag \\
&&~~~~\frac{(\ell^{8}-180\ell^{6}N^{2}+5400\ell^{4}N^{4}-162000%
\ell^{2}N^{6}+810000N^{8}) }{N^{4}\ell^{2}\sqrt{\ell^{4}+900N^{4}-180N^{2}%
\ell^{2}}}\Bigg)\Bigg]  \notag
\end{eqnarray}

Since $r_{b+}$ diverges as $\ell \rightarrow \infty $, we must consider only
the $r_{b-}$ branch in the flat space limit. We find $r_{b-}$ $\rightarrow
r_{0}=3N$ and
\begin{eqnarray*}
I_{Bolt}(r_{b}=r_{b-}) &\rightarrow &32\pi ^{2}N^{4} \\
S_{Bolt}(r_{b}=r_{b-}) &\rightarrow &96\pi ^{2}N^{4} \\
C_{Bolt}(r_{b}=r_{b-}) &\rightarrow &-384\pi ^{2}N^{4}
\end{eqnarray*}
and the Bolt mass parameter (\ref{6dmb}) is
\begin{equation}
m_{b}\rightarrow \frac{(r_{b}^{4}-6N^{2}r_{b}^{2}-3N^{4})}{6r_{b}}=\frac{4}{3%
}N^{3}  \label{6dmbflat}
\end{equation}
confirming the results of ref. \cite{AC}.

\begin{figure}[tbp]
\centering
\begin{minipage}[c]{.55\textwidth}
         \centering
         \includegraphics[width=\textwidth]{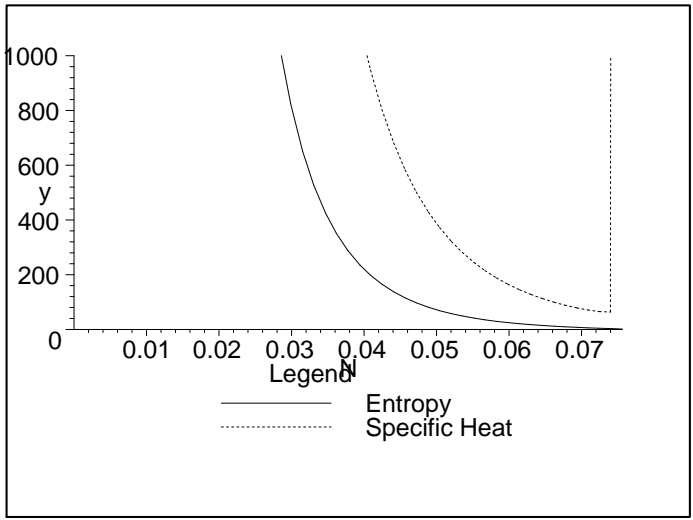}
         \caption{Plot of Entropy and Specific Heat vs. $N$ for the
upper branch Bolt solutions in 6 dimensions.}
         \label{6dSCBoltplot}
     \end{minipage}
\end{figure}

\begin{figure}[tbp]
\centering
\begin{minipage}[c]{.45\textwidth}
         \centering
         \includegraphics[width=\textwidth]{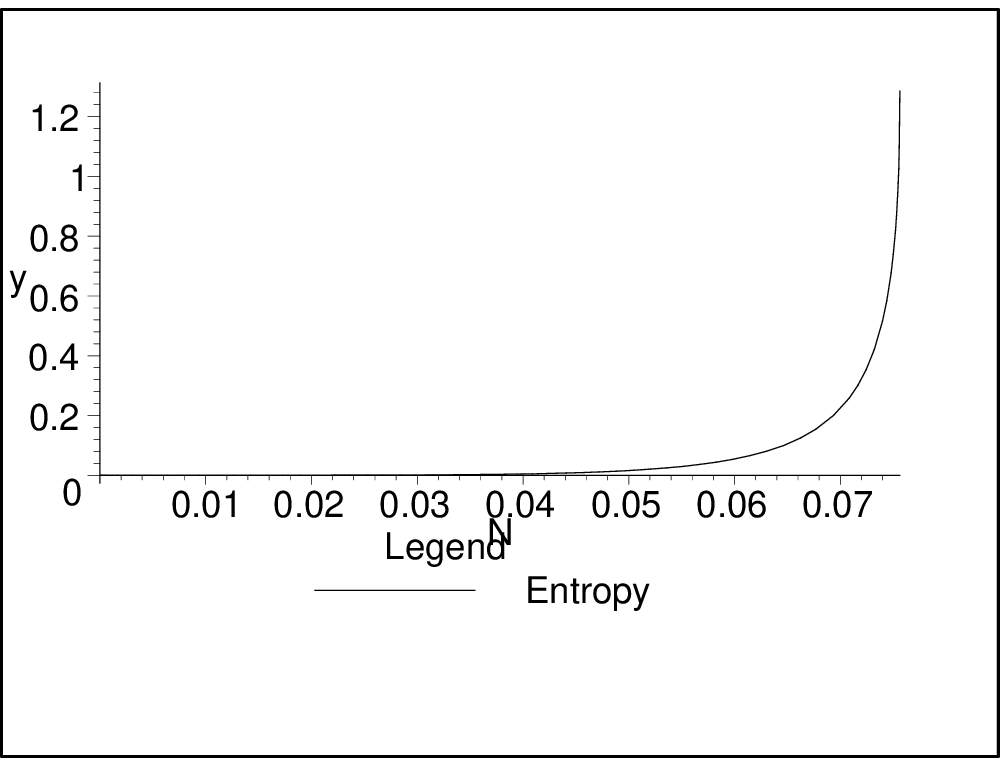}
         \caption{Plot of lower branch Bolt Entropy vs. $N$ in 6 dimensions}
         \label{6dSBmPlot}
     \end{minipage}
\begin{minipage}[c]{.35\textwidth}
     \end{minipage}
\begin{minipage}[c]{.45\textwidth}
         \centering
         \includegraphics[width=\textwidth]{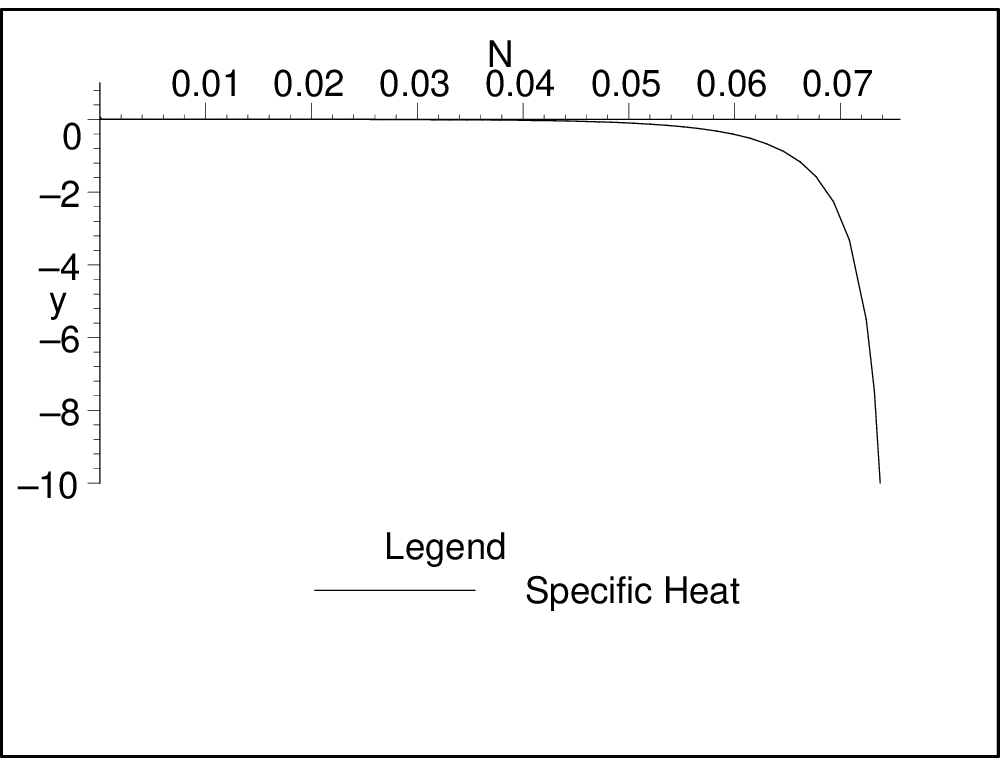}
         \caption{Plot of lower branch Bolt Specific Heat vs. $N$ in 6
dimensions}
         \label{6dCBmPlot}
     \end{minipage}
\end{figure}

As with the NUT case, we again require that the entropy and specific heat
both be positive for a thermodynamically stable solution. The upper branches
for the entropy and specific heat are both positive (for $0<N<N_{\text{max}}$%
, see Figure \ref{6dSCBoltplot}), and thus the upper branch solution is
thermodynamically stable. However, although the lower branch entropy is
positive (in this range), the lower branch specific heat (Figures \ref%
{6dSBmPlot}, \ref{6dCBmPlot}) is negative, giving a thermodynamically
unstable lower branch solution in 6 dimensions.

\subsection{N\"other conserved quantities}

Consider the Taub-bolt solution (\ref{6dmetric1}) with mass parameter given
by eq. (\ref{6dmb}) and identification period by eq. (\ref{6dbetaBolt}). We
shall use the Taub-NUT metric (\ref{6dmetric1}) as the background, with mass
parameter and identification period respectively given by (\ref{mn6d}, \ref%
{betagen}). Note that, despite the name, the mass parameters $m_{b}$ and $%
m_{n}$ have \textit{a priori} nothing to do with the mass of the solutions
(except perhaps via some Newtonian approximation, whose meaning is not clear
in this context). We shall shortly see that the N\"{o}ther theorem furnishes
us with justification for these names.

Let us then choose a spacetime vector field
\begin{equation}
\xi =\partial _{\tau }+a\partial _{\theta _{1}}+b\partial _{\theta _{2}}
\end{equation}
which by definition produces the N\"{o}ther conserved quantity $
Q=m+aJ_{1}+bJ_{2}$ of the corresponding dynamical metric relative to the
background. We evaluate the superpotential (\ref{superpotential}) on the
solutions $(g,\bar{g})$ and then integrating on a spatial region $(\tau
=\tau _{0},r=r_{0})$, expanding in Taylor series around $r_0=\infty $. By
keeping the three contributions of the superpotential separate, we obtain $
Q=Q_{1}+ Q_{2}+ Q_{3}$, where we set:
\begin{eqnarray*}
Q_{1} &=&\hbox{$4\pi^2 \over \ka
\ell^2$}\Big(2r_0^{5}-4N^{2}r_0^{3}- \hbox{$2r_0\over
3$}N^{2}(21N^{2}-4\ell^{2})+\hbox{$1\over  r_b $} (3r_{b}^{6}-(15N^{2}-%
\ell^{2})r_{b}^{4} \\
&&+3N^{2}(15N^{2}-2\ell^{2})r_{b}^{2}+3N^{4}(5N^{2}-\ell^{2}))\Big) %
+O(r^{-1}_0) \\
&& \\
Q_{2} &=&\hbox{$4\pi^2\over 3\ka r_b
\ell^2$}\Big(3r_{b}^{6}-(15N^{2}-\ell^{2})r_{b}^{4}+3N^{2}(15N^{2}
-2\ell^{2})r_{b}^{2}+ \\
&&-8N^{3}(6N^{2}-\ell^{2})r_{b}+3N^{4}(5N^{2}-\ell^{2})\Big)+O(r^{-1}_0) \\
&& \\
Q_{3} &=&\hbox{$4\pi^2\over \ka
\ell^2$}\Big(-2r_0^{5}+4N^{2}r_0^{3}+\hbox{$2 r_0\over
3$}N^{2}(21N^{2}-4\ell^{2})-8N^{3}(6N^{2}-\ell^{2})\Big)+O(r_0^{-1})
\end{eqnarray*}
Although $Q_{1}$ and $Q_{3}$ diverge as $r_0 \rightarrow \infty $, the total
conserved quantity $Q$ does not. In fact
\begin{eqnarray*}
Q &=&\hbox{$16\pi^2\over \ka \ell^2$}\Big(r_{b}^{5}-\hbox{$1\over
3$}(15N^{2}-\ell^{2})r_{b}^{3}+N^{2}(15N^{2}-2\ell^{2})r_{b}-
\hbox{$8\over
3$}N^{3}(6N^{2}-\ell^{2})+ \\
&&+\hbox{$1\over  r_b $}N^{4}(5N^{2}-\ell^{2})\Big)+O(r_0^{-1})
\end{eqnarray*}
No contribution from the angular parts of $\xi $ survives, hence $J_{1}=0$
and $J_{2}=0$. According to the interpretation of $Q$ as relative mass, we
have
\begin{equation}
Q=\hbox{$32\pi^2\over\ka$}(m_{b}-m_{n})  \label{Qmass6}
\end{equation}

The first law of thermodynamics then gives
\begin{equation}
\delta S=12\pi N\delta Q  \label{firstlaw6}
\end{equation}%
which can be interpreted as the relative entropy of the two solutions. We
obtain

\begin{eqnarray*}
&&S=\hbox{$16\pi^3\over 3375 \ka  N^3\ell^2$}\Big(r_{b}(\ell^{8}
-90N^{2}\ell^{6}-300N^{4}\ell^{4}-27000N^{6}\ell^{2}+540000N^{8})+ \\
&&\qquad
+N(30N^{2}-\ell^{2})(3\ell^{6}+80N^{2}\ell^{4}+1500N^{4}\ell^{2}-18000N^{6})%
\Big)
\end{eqnarray*}
This is different from the relative entropy $S_{Bolt}-S_{NUT}$ given from
eqs. (\ref{6dSB},\ref{6dSN}) However, the difference is an integration
constant
\begin{equation}
S=\hbox{$4\pi\over\ka$}\Big(S_{Bolt}-S_{NUT}-{\frac{112\pi^{2}}{675}}l^{4} %
\Big)
\end{equation}
The two methods provide the same classical entropy, up an integration
constant as discussed earlier.

\section{Eight Dimensional Solutions}

As in six dimensions, there are again two possible forms for the metric (%
\cite{AC}). The $U(1)$ fibration over $S^{2}\times S^{2}\times S^{2} $ gives
the following form;
\begin{eqnarray}
ds^{2} &=&F(r)(d\tau +2N\cos (\theta _{1})d\phi _{1}+2N\cos (\theta
_{2})d\phi _{2}+2N\cos (\theta _{3})d\phi _{3})^{2}+\frac{dr^{2}}{F(r)}
+(r^{2}-N^{2})\left( d\theta _{1}^{2}\right.  \notag \\
&&\left. +\sin ^{2}(\theta _{1})d\phi _{1}^{2}+d\theta _{2}^{2}+\sin
^{2}(\theta _{2})d\phi _{2}^{2}+d\theta _{3}^{2}+\sin ^{2}(\theta
_{3})d\phi_{3}^{2}\right)  \label{8dmetric}
\end{eqnarray}
There is also a second form using the base space $\mathcal{B} = S^2 \times
\mathbb{CP}^2$, given by
\begin{equation}
ds^2 = F(r) (d\tau + A )^2 + \frac{dr^2}{F(r)} + (r^2-N^2)\left(
d\Sigma_2^{~2} + d\Omega_2^{~2} \right)
\end{equation}
where $d\Sigma_2^{~2}$ is again the $\mathbb{CP}^2$ form (\ref{CP2Metric}),
and $d\Omega_2^{~2}$ is the metric of the 2-sphere $S^2$;
\begin{equation}
d\Omega_2^{~2} = d\theta_1^{~2} + \sin^2 (\theta_1) d\phi_1^{~2}
\label{S2Metric}
\end{equation}
$A$ is given in eight dimensions by
\begin{equation}
A = 2N \cos (\theta_1) d\phi_1 + \frac{u^2 N}{2\left( 1 + \frac{u^2}{6}
\right)} \left( d\psi + \cos (\theta) d\phi \right)
\end{equation}
(where $\theta_1$, $\phi_1$ are the coordinates for the 2-sphere). The form
of $F(r)$ is again the same for either choice of metric;
\begin{equation}
F(r)=\frac{5r^{8}+(\ell ^{2}-28N^{2})r^{6}+5N^{2}(14N^{2}-\ell
^{2})r^{4}+5N^{4}(3\ell ^{2}-28N^{2})r^{2}-10mr\ell ^{2}+5N^{6}(\ell
^{2}-7N^{2})}{5\ell ^{2}(r^{2}-N^{2})^{3}}  \label{8dFr}
\end{equation}
Also, as in six dimensions, these metrics give exactly the same results,
with the only difference in the calculation method being the volume element
used.

The action (\ref{action}) before specializing to either the NUT or Bolt case
is
\begin{equation}
I=\frac{8\beta \pi ^{2}(5m\ell
^{2}-35N^{4}r_{+}^{3}-5r_{+}^{7}+35N^{6}r_{+}+21N^{2}r_{+}^{5})}{5\ell ^{2}}
\label{8dItot}
\end{equation}
where the period is now
\begin{equation}
\beta =16\pi N  \label{8dbeta}
\end{equation}
In the flat space limit, again using $r_{b-}$ as the largest positive root
of $F(r)$, (\ref{8dItot}) becomes
\begin{equation}
I\rightarrow 8m\pi ^{2}\beta  \label{8dItotflat}
\end{equation}
We also obtain
\begin{equation}
\mathfrak{M}=48\pi ^{2}m  \label{8dM}
\end{equation}
for the conserved mass in 8 dimensions.

\subsection{Taub-NUT-AdS Solution}

\begin{figure}[tbp]
\centering
\begin{minipage}[c]{.45\textwidth}
         \centering
         \includegraphics[width=\textwidth]{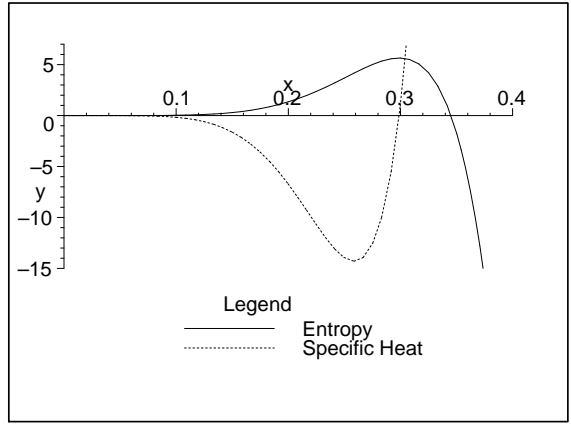}
         \caption{Plot of NUT Entropy and Specific Heat vs. $N$ (8-dimensions)}
         \label{8dSCPlot}
     \end{minipage}
\end{figure}

The mass parameter for the eight dimensional NUT solution with $r_{+}=N$ is
fixed to be
\begin{equation}
m_{n}=\frac{8N^{5}(\ell ^{2}-8N^{2})}{5\ell ^{2}}  \label{8dmn}
\end{equation}
Notice this has the same overall sign in the large-$\ell /N$ limits as the
four dimensional case. The action for the NUT is
\begin{equation}
I_{NUT}=\frac{1024\pi ^{3}N^{6}(-6N^{2}+\ell ^{2})}{5\ell ^{2}}  \label{8dIN}
\end{equation}
and the entropy (which again satisfies the first law) and specific heat are
\begin{eqnarray}
S_{NUT} &=&\frac{1024\pi ^{3}N^{6}(5\ell ^{2}-42N^{2})}{5\ell ^{2}}
\label{8dSN} \\
C_{NUT} &=&\frac{6144\pi ^{3}N^{6}(-5\ell ^{2}+56N^{2})}{5\ell ^{2}}
\label{8dCN}
\end{eqnarray}
Note here that the overall signs of the above quantities are the same as the
four dimensional quantities, as noted above. In eight dimensions, the action
becomes negative for $N>{\frac{\ell }{\sqrt{6}}}$. Now, though, the entropy
is negative for $N>\ell \sqrt{{\frac{5}{42}}}$, and the specific heat for $%
N<\ell \sqrt{{\frac{5}{56}}}$. Hence in eight dimensions there is a range of
$N$ for which the entropy and specific heat are both positive, as can be
seen in Figure \ref{8dSCPlot};
\begin{equation}
\ell \sqrt{\frac{5}{56}}<N<\ell \sqrt{\frac{5}{42}}
\end{equation}
and thermally stable solutions exist within this range of parameters. The
high temperature limits for the action, specific heat and entropy are all
still zero.

In the flat space limit, (\ref{8dIN}), (\ref{6dSN}) and (\ref{8dCN}) are
\begin{eqnarray*}
I_{NUT} &\rightarrow &\frac{1024}{5}\pi ^{3}N^{6} \\
S_{NUT} &\rightarrow &1024\pi ^{3}N^{6} \\
C_{NUT} &\rightarrow &-6144\pi ^{3}N^{6}
\end{eqnarray*}
indicating that asymptotically locally flat pure NUT solutions in 8
dimensions are thermally unstable.

\subsection{Taub-Bolt-AdS Solution}

We obtain
\begin{equation}
m_{b}=\frac{5r_{b}^{8}+(\ell ^{2}-28N^{2})r_{b}^{6}+5N^{2}(14N^{2}-\ell
^{2})r_{b}^{4}+5N^{4}(3\ell ^{2}-28N^{2})r_{b}^{2}+5N^{6}(\ell ^{2}-7N^{2})}{
10\ell ^{2}r_{b}}  \label{8dmb}
\end{equation}
for the mass parameter in terms of the bolt radius. Regularity implies that $%
F^{\prime }(r_{b})={\frac{1}{4N}}$, which has two solutions
\begin{equation}
r_{b\pm }=\frac{\ell ^{2}\pm \sqrt{\ell ^{4}-448N^{2}\ell ^{2}+3136N^{4}}}{
56N}  \label{8drbpm}
\end{equation}
each of which must be real, with $r_{b}>N$. This implies
\begin{equation}
N\leq \frac{\ell }{28}\left( \sqrt{35}-\sqrt{21}\right) =N_{\text{max}}
\label{Nboltlim8}
\end{equation}

The Bolt action in this case is
\begin{equation}
I_{Bolt}=\frac{16\pi ^{3}(5r_{b}^{8}-(\ell
^{2}+14N^{2})r_{b}^{6}+5N^{2}r_{b}^{4}\ell ^{2}-5N^{4}(3\ell
^{2}-14N^{2})r_{b}^{2}-5N^{6}(\ell ^{2}-7N^{2}))}{5(7N^{2}-7r_{b}^{2}-\ell
^{2})}  \label{8dIB}
\end{equation}
where now
\begin{equation}
\beta _{Bolt}=\frac{10\pi \ell ^{2}(r_{b}^{2}-N^{2})^{4}}{\rho }
\label{8dbetaBolt}
\end{equation}
with
\begin{eqnarray*}
\rho &=&5r_{b}^{9}-20r_{b}^{7}N^{2}+2N^{2}(\ell
^{2}+7N^{2})r_{b}^{5}+20N^{4}(7N^{2}-\ell ^{2})r_{b}^{3}+25m\ell
^{2}r_{b}^{2} \\
&&+5N^{6}(49N^{2}-6\ell ^{2})r_{b}+5m\ell ^{2}N^{2}
\end{eqnarray*}
The Bolt entropy is
\begin{eqnarray}
S_{Bolt} &=&\frac{16\pi ^{3}}{5(7r_{b}^{2}+\ell ^{2}-7N^{2})}\left(
35r_{b}^{8}+(5\ell ^{2}-182N^{2})r_{b}^{6}-5N^{2}(5\ell
^{2}-84N^{2})r_{b}^{4}\right.  \notag \\
&&\left. ~~~~~~~~~~~~~~~~~~~~~~~-5N^{4}(154N^{2}-15\ell
^{2})r_{b}^{2}+25N^{6}(\ell ^{2}-7N^{2})\right)  \label{Sbolt8}
\end{eqnarray}
and the specific heat is
\begin{eqnarray}
C_{Bolt}(r_{b}=r_{b\pm }) &=&\frac{3\pi ^{3}}{150590720N^{6}}\Bigg[\ell
^{8}(190512N^{4}-2240N^{2}\ell ^{2}+5\ell ^{4})  \notag \\
&&\pm \frac{1}{\ell ^{2}\sqrt{\ell ^{4}+3136N^{4}-448N^{2}\ell ^{2}}}\Big( %
193434623148032N^{16}  \notag \\
&&-31087707291648N^{14}\ell ^{2}+771024486400N^{12}\ell ^{4}  \notag \\
&&+8260976640N^{10}\ell ^{6}+36879360N^{8}\ell ^{8}-16332288N^{6}\ell ^{10}
\notag \\
&&+574672N^{4}\ell ^{12}-3360N^{2}\ell ^{14}+5\ell ^{16}\Big)\Bigg]
\label{Cbolt8}
\end{eqnarray}

\begin{figure}[tbp]
\centering
\begin{minipage}[c]{.55\textwidth}
         \centering
         \includegraphics[width=\textwidth]{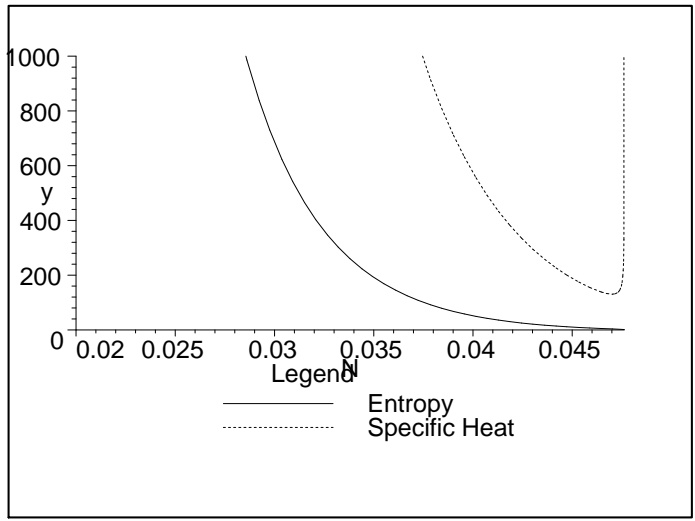}
         \caption{Plot of Entropy and Specific Heat vs. $N$ for the 8
dimensional upper branch Bolt solutions.}
         \label{8dSCBoltplot}
     \end{minipage}
\end{figure}

\begin{figure}[tbp]
\centering
\begin{minipage}[c]{.45\textwidth}
         \centering
         \includegraphics[width=\textwidth]{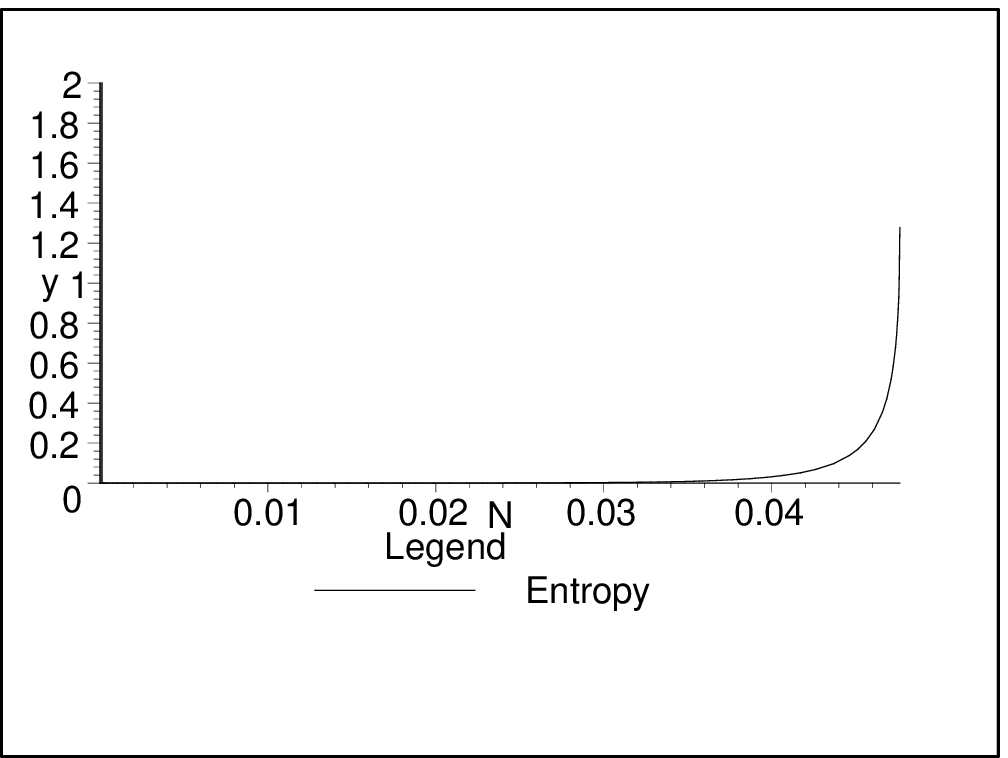}
         \caption{Plot of lower branch Bolt Entropy vs. $N$ (8 dimensions)}
         \label{8dSBmPlot}
     \end{minipage}
\begin{minipage}[c]{.40\textwidth}
     \end{minipage}
\begin{minipage}[c]{.45\textwidth}
         \centering
         \includegraphics[width=\textwidth]{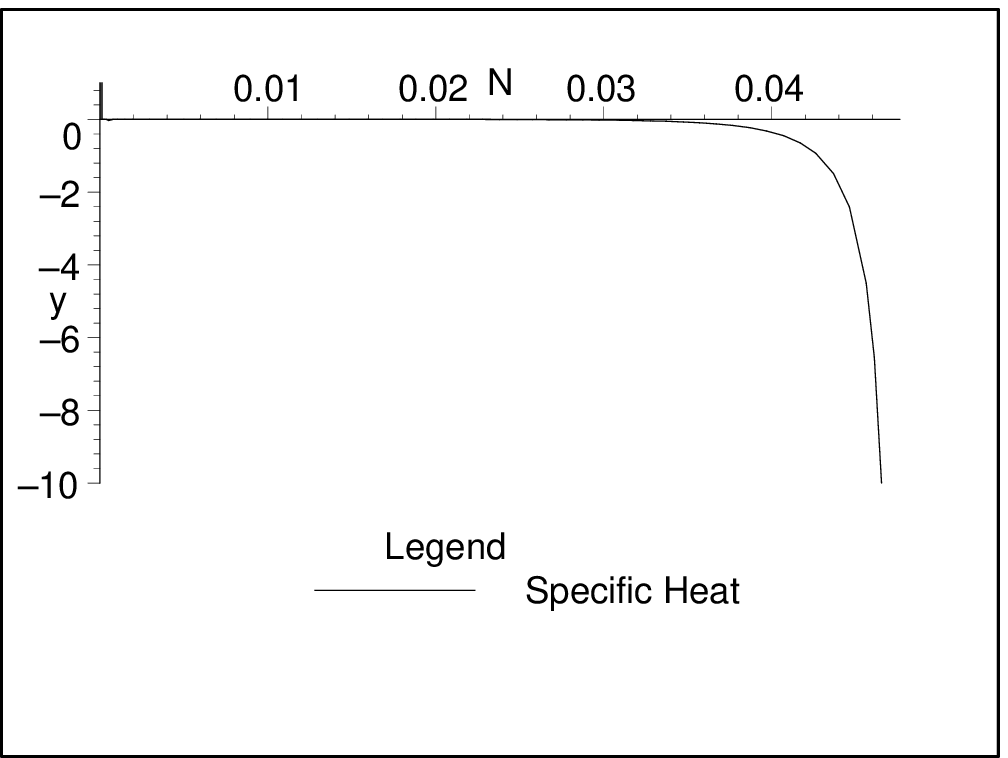}
         \caption{Plot of lower branch Bolt Specific Heat vs. $N$ (8
dimensions)}
         \label{8dCBmPlot}
     \end{minipage}
\end{figure}

As in six dimensions, the upper branch Bolt quantities are all ($\pm $)
infinity in the high temperature limit; the lower branch quantities are all
zero. We find (figure \ref{8dSCBoltplot}) that the upper branch solutions
for both the entropy and specific heat are again thermodynamically stable
for $0<N<N_{\text{max}}$, and the lower branch solutions are again unstable,
as can be seen in Figures \ref{8dSBmPlot}, \ref{8dCBmPlot}.

The flat space limits for the lower branch solutions are,
\begin{eqnarray}
I_{Bolt}(r_{b}=r_{b-}) &\rightarrow &\frac{48976}{5}N^{6}\pi ^{3}  \notag \\
S_{Bolt}(r_{b}=r_{b-}) &\rightarrow &48976N^{6}\pi ^{3}  \label{ALF8bolt} \\
C_{Bolt}(r_{b}=r_{b-}) &\rightarrow &-293856N^{6}\pi ^{3}  \notag
\end{eqnarray}
where the lower branch Bolt radius now goes to $4N$. Hence in the flat space
limit, the Bolt mass becomes
\begin{equation}
m_{b}\rightarrow \frac{r_{b}^{6}-5N^{2}r_{b}^{4}+15N^{4}r_{b}^{2}+5N^{6}}{
10r_{b}}=\frac{3061}{40}N^{5}  \label{MALF8bolt}
\end{equation}
These solutions have negative specific heat, and so are unstable for all
values of the parameters.

\subsection{N\"other conserved quantities}

We now consider the Taub-bolt solution (\ref{8dmetric}) as the dynamical
metric with mass parameter and identification period given by (\ref{8dmb}, %
\ref{8dbetaBolt}). The Taub-NUT metric (\ref{8dmetric}) with mass parameter (%
\ref{8dmn}) will be taken to be the background. Choosing the spacetime
vector field
\begin{equation}
\xi =\partial _{\tau }+a\partial _{\theta _{1}}+b\partial _{\theta
_{2}}+c\partial _{\theta _{3}}
\end{equation}%
we evaluate the superpotential (\ref{superpotential}) on the solutions $(g,
\bar{g})$ then integrating on a spatial region $(\tau =\tau _{0},r=r_{0})$
as before, taking the limit $r_{0}\rightarrow \infty $. By keeping the three
contributions of the superpotential separate, we obtain $Q=Q_{1} + Q_{2} +
Q_{3}$, where we set:
\begin{eqnarray*}
Q_{1} &=&{\frac{16\pi^{3}}{\kappa \ell^{2}}}\Big(2r_0^{7}-6N^{2}r_0^{5}-
\hbox{$2\over
5$}N^{2}(N^{2}-2\ell ^{2})r_0^{3}+\hbox{$2\over 5$} N^{4}(139N^{2}-18\ell
^{2})r_0 \\
&&+(5r_{b}^{7}-(28N^{2}-\ell ^{2})r_{b}^{5}+5N^{2}(14N^{2}-\ell
^{2})r_{b}^{3}-5N^{4}(28N^{2}-3\ell ^{2})r_{b} \\
&&-\hbox{$5\over r_b$}N^{6}(7N^{2}-\ell ^{2})\Big)+O(r_0^{-1}) \\
&& \\
Q_{2} &=&\hbox{$16\pi^3\over 5\ka
\ell^2$}\Big(5r_{b}^{7}-(28N^{2}-\ell^{2})r_{b}^{5}+5N^{2}(14N^{2}
-\ell^{2})r_{b}^{3}-5N^{4}(28N^{2}-3\ell^{2})r_{b} \\
&&+16N^{5}(8N^{2}-\ell^{2})-\hbox{$5\over
r_b$}N^{6}(7N^{2}-\ell^{2})\Big) +O(r_0^{-1}) \\
&& \\
Q_{3} &=&\hbox{$16\pi^3\over \ka
\ell^2$}\Big(-2r_0^{7}+6N^{2}r_0^{5}+ \hbox{$2\over
5$}N^{2}(N^{2}-2\ell^{2})r_0^{3}-\hbox{$2\over
5$}N^{4}(139N^{2}-18\ell^{2})r_0 \\
&&+16N^{5}(8N^{2}-\ell^{2})\Big)+O(r_0^{-1}) \\
&&
\end{eqnarray*}

Again, $Q_{1}$ and $Q_{3}$ diverge as $r_0 \rightarrow \infty $, while the
total conserved quantity $Q$ does not. In fact
\begin{eqnarray*}
Q &=&{\frac{96\pi^{3}}{\kappa \ell^{2}}}\Big(r_{b}^{7}-\hbox{$1\over
5$}(28N^{2}-\ell^{2})r_{b}^{5}+N^{2}(14N^{2}-
\ell^{2})r_{b}^{3}-N^{4}(28N^{2}-3\ell ^{2})r_{b}+ \\
&&-\hbox{$1\over r_b$}N^{6}(7N^{2}-\ell^{2})+\hbox{$16\over
5$}N^{5}(8N^{2}-\ell^{2})\Big)+O(r_0^{-1})
\end{eqnarray*}

No contribution form the angular part of $\xi $ survives, hence $
J_{1}=J_{2}=J_{3}=0$. We find that
\begin{equation}
Q=\hbox{$192\pi^3\over \ka$}(m_{b}-m_{n})  \label{Q8}
\end{equation}
Again, from the first law of thermodynamics
\begin{equation}
\delta S=16\pi N\delta Q  \label{firstlaw8}
\end{equation}
we can obtain $S$ as the relative entropy of the two solutions. Explicitly,

\begin{eqnarray*}
S &=&\hbox{$ \pi^4 \over 1344560\ka N^5 \ell^2$}\Big((5\ell^{12} -
2240N^{2}\ell^{10}+187376N^{4}\ell^{8}+702464N^{6}\ell^{6} \\
&&+46713856\ell^{4}N^{8}+2202927104N^{10}\ell^{2} -46261469184N^{12})r_{b}+
\\
&&-4N(+5\ell^{12}-1715N^{2}\ell^{10}+3136N^{4}\ell^{8}
+208544N^{6}\ell^{6}+9834496N^{8}\ell^{4} \\
&&+1170305024N^{10}\ell^{2}-11565367296N^{12})\Big) \\
&=&\hbox{$4\pi\over
\kappa$}\Big((S_{Bolt}-S_{NUT})+{\frac{17\pi^{3}}{343}} \ell^{6}\Big)
\end{eqnarray*}

\section{Ten Dimensional Solution}

In ten dimensions, there are three possible forms for the metric (\cite{AC}%
). A $U(1)$ fibration over $S^{2}\times S^{2}\times S^{2}\times S^{2}$ gives
the following form
\begin{eqnarray}
ds^{2} &=&F(r)(d\tau +2N\cos (\theta _{1})d\phi _{1}+2N\cos (\theta
_{2})d\phi _{2}+2N\cos (\theta _{3})d\phi _{3}+2N\cos (\theta _{4})d\phi
_{4})^{2}  \notag \\
&&+\frac{dr^{2}}{F(r)}+(r^{2}-N^{2})\left( d\theta _{1}^{2}+\sin ^{2}(\theta
_{1})d\phi _{1}^{2}+d\theta _{2}^{2}+\sin ^{2}(\theta _{2})d\phi
_{2}^{2}\right.  \notag \\
&&\left. +d\theta _{3}^{2}+\sin ^{2}(\theta _{3})d\phi _{3}^{2}+d\theta
_{4}^{2}+\sin ^{2}(\theta _{4})d\phi _{4}^{2}\right)  \label{10dmetric1}
\end{eqnarray}
A $U(1)$ fibration over $S^{2}\times S^{2}\times \mathbb{CP}^{2}$ can also
be used, giving a metric with the form
\begin{equation}
ds^{2}=F(r)(d\tau +A)^{2}+\frac{dr^{2}}{F(r)}+(r^{2}-N^{2})\left( d\Sigma
_{2}^{~2}+d\Omega _{2}^{~2}+d\Omega _{2}^{^{\prime }~2}\right)
\end{equation}
where $d\Sigma _{2}^{~2}$ is again the metric over $\mathbb{CP}^{2}$ (\ref%
{CP2Metric}), and the $d\Omega _{2}^{~2}$, $d\Omega _{2}^{^{\prime }~2}$ are
the metrics over the two 2-spheres (\ref{S2Metric}). For this fibration the
one-form $A$ is given by
\begin{equation}
A=2N\cos (\theta _{1})d\phi _{1}+2N\cos (\theta _{2})d\phi _{2}+\frac{u^{2}N
}{2\left( 1+\frac{u^{2}}{6}\right) }\left( d\psi +\cos (\theta )d\psi \right)
\end{equation}
(where $\theta _{1}$, $\phi _{1}$ are the coordinates for the first sphere, $%
\theta _{2}$, $\phi _{2}$ for the second, and $\theta $, $\phi $ are the
coordinates for $\mathbb{CP}^{2}$).

A third form of the metric in ten dimensions can be calculated using a $U(1)$
fibration over $\mathbb{CP}^{2}\times \mathbb{CP}^{2}$:
\begin{equation}
ds^{2}=F(r)(d\tau +A)^{2}+\frac{dr^{2}}{F(r)}+(r^{2}-N^{2})\left( d\Sigma
_{2}^{~2}+d\Sigma _{2}^{^{\prime }~2}\right)
\end{equation}
with $d\Sigma _{2}^{^{\prime }~2}$ being another metric over $\mathbb{CP}%
^{2} $, and $A$ given by
\begin{equation}
A=\frac{u^{\prime }{}^{2}N}{2\left( 1+\frac{u^{\prime }{}^{2}}{6}\right) }
\left( d\psi ^{\prime }+\cos (\theta ^{\prime })d\phi ^{\prime }\right) +
\frac{u^{2}N}{2\left( 1+\frac{u^{2}}{6}\right) }\left( d\psi +\cos (\theta
)d\phi \right)
\end{equation}
In all of these metrics, the form of $F(r)$ in ten dimensions is given by
\begin{eqnarray}
F(r) &=&\frac{1}{35\ell ^{2}(r^{2}-N^{2})^{4}}\Big[35r^{10}+5(\ell
^{2}-45N^{2})r^{8}+14N^{2}(45N^{2}-2\ell ^{2})r^{6}  \notag \\
&&+70N^{4}(\ell ^{2}-15N^{2})r^{4}+35N^{6}(45N^{2}-4\ell ^{2})r^{2}-70mr\ell
^{2}+35N^{8}(9N^{2}-\ell ^{2})\Big]~~  \label{Fr10}
\end{eqnarray}
All three metrics will give the same results for the action and other
quantities to be calculated, apart from the overall volume factors from the
compact spaces.

The total action before specifying to the NUT or Bolt case is
\begin{equation}
I=\frac{-32\pi ^{3}\beta
(35r_{+}^{9}-180N^{2}r_{+}^{7}+378N^{4}r_{+}^{5}-420N^{6}r_{+}^{3}+315N^{8}r_{+}-35m\ell ^{2})
}{35\ell ^{2}}  \label{Itot10d}
\end{equation}
Counter-terms for $I_{ct}$ have only been computed up to 9 dimensions \cite%
{DasMann}. Rather than extend this expression to 10 dimensions, we have
found (see section \ref{sec:Gensol}) that for these class of metrics, only
the first term in the counter-term expansion (\ref{Lagrangianct}) is
necessary to compute the finite action; the remaining terms of the expansion
will only cancel divergences, making no contribution to the finite part.

The period in 10 dimensions is given by
\begin{equation}
\beta =20\pi N
\end{equation}
The conserved mass, found from the general expression (\ref{MassGen})
(again, see section \ref{sec:Gensol}) is
\begin{equation}
\mathfrak{M}=256\pi ^{3}m
\end{equation}
The above expression for the conserved mass can be shown to satisfy the
first law of thermodynamics for black holes in both the NUT and Bolt cases.

In the flat space limit, again substituting the root $r_{b-}$ (\ref{10drbpm}%
), (\ref{Itot10d}) becomes
\begin{equation*}
I\rightarrow 32m\pi ^{3}\beta
\end{equation*}

\subsection{Taub-NUT-AdS Solution}

The mass parameter in 10 dimensions is
\begin{equation}
m_{n}=\frac{64N^{7}(10N^{2}-\ell ^{2})}{35\ell ^{2}}  \label{mn10d}
\end{equation}
upon elimination of conical singularities at $r=N$. Using the methods of
sec. \ref{sec:Gensol}, the action is
\begin{equation}
I_{NUT}=\frac{8192\pi ^{4}N^{8}(8N^{2}-\ell ^{2})}{7\ell ^{2}}  \label{In10}
\end{equation}
and the entropy and specific heat are
\begin{eqnarray}
S_{NUT} &=&\frac{8192\pi ^{4}N^{8}(72N^{2}-7\ell ^{2})}{7\ell ^{2}}
\label{Snut10} \\
C_{NUT} &=&\frac{65536\pi ^{4}N^{8}(7\ell ^{2}-90N^{2})}{7\ell ^{2}}
\label{Cnut10}
\end{eqnarray}
The overall signs here are the same as in the 6 dimensional NUT case, as
expected. Note that the action becomes negative for $N<{\frac{\ell }{\sqrt{8}
}}$, the entropy for $N<\ell \sqrt{{\frac{7}{72}}}$, and the specific heat
for $N>\ell \sqrt{{\frac{7}{90}}}$. A plot of the entropy and the specific
heat as functions of $N/\ell $ (Figure \ref{6dSCPlot}) is similar to the 6
dimensional case: there is no region in which both are simultaneously
positive, and so the 10 dimensional Taub-NUT-AdS solution is
thermodynamically unstable. As before the high temperature limits of the
entropy and specific heat are zero, as is the high temperature limit of the
action.

The flat space limits ($\ell \rightarrow \infty $) of these quantities in 10
dimensions are
\begin{eqnarray*}
I_{NUT} &\rightarrow &-\frac{8192\pi ^{4}N^{8}}{7} \\
S_{NUT} &\rightarrow &-8192\pi ^{4}N^{8} \\
C_{NUT} &\rightarrow &65536\pi ^{4}N^{8}
\end{eqnarray*}%
indicating that asymptotically locally flat spacetimes with NUT charge in
ten dimensions are thermodynamically unstable for all values of $N$. \

\subsection{Taub-Bolt-AdS Solution}

Setting $r_{+}=r_{b}$, the mass parameter in 10 dimensions is
\begin{eqnarray}
m_{b} &=&\frac{1}{70l^{2}r_{b}}\Big[35r_{b}^{10}+5(\ell
^{2}-45N^{2})r_{b}^{8}+14N^{2}(45N^{2}-2\ell ^{2})r_{b}^{6}  \notag \\
&&+70N^{4}(\ell ^{2}-15N^{2})r_{b}^{4}+35N^{6}(45N^{2}-4\ell
^{2})r_{b}^{2}+35N^{8}(9N^{2}-\ell ^{2})\Big]  \label{10dmb}
\end{eqnarray}
Condition (ii) now becomes $F^{\prime }(r_{b})={\frac{1}{5N}}$, forcing $%
r_{b}$ to be
\begin{equation}
r_{b\pm }=\frac{\ell ^{2}\pm \sqrt{\ell ^{4}-900N^{2}\ell ^{2}+8100N^{4}}}{
90N}  \label{10drbpm}
\end{equation}
As before, we require this to be real, and $r_{b}>N$, which gives
\begin{equation}
N\leq \frac{\sqrt{30}-2\sqrt{5}}{30}=N_{\text{max}}  \label{Nmax10}
\end{equation}
Employing the methods of section \ref{sec:Gensol} the Bolt action becomes
\begin{eqnarray}
I_{Bolt} &=&\frac{-64\pi ^{4}}{35(9r_{b}^{2}-9N^{2}+\ell ^{2})}\Big[ %
35r_{b}^{10}-5(\ell ^{2}+27N^{2})r_{b}^{8}+14N^{2}(2\ell
^{2}+9N^{2})r_{b}^{6}  \notag \\
&&-70N^{4}(\ell ^{2}-3N^{2})r_{b}^{4}-35N^{6}(27N^{2}-4\ell
^{2})r_{b}^{2}-35N^{8}(9N^{2}-\ell ^{2})\Big]  \label{10dIB}
\end{eqnarray}
Regularity implies ${\beta =}\left| {\frac{4\pi }{F^{\prime }(r_{b})}}
\right| $, which gives
\begin{equation}
\beta _{Bolt}=\frac{70\pi (r_{b}^{2}-N^{2})^{5}\ell ^{2}}{\rho }
\label{10dbetaBolt}
\end{equation}
with
\begin{eqnarray}
\rho &=&35r_{b}^{11}-175r_{b}^{9}N^{2}+2N^{2}(4\ell
^{2}+135N^{2})r_{b}^{7}+14N^{4}(15N^{2}-4\ell ^{2})r_{b}^{5}  \notag \\
&&+35N^{6}(8\ell ^{2}-75N^{2})r_{b}^{3}+245mr_{b}^{2}\ell ^{2}+35N^{8}(8\ell
^{2}-81N^{2})r_{b}+35m\ell ^{2}N^{2}  \notag
\end{eqnarray}
We again note that by substituting in $r_{b}=r_{b-}$ and $m=m_{b}$ into $%
\beta _{Bolt}$, the temperature is the same as in the NUT case. The Bolt
entropy is
\begin{eqnarray}
S_{Bolt} &=&\frac{64\pi ^{4}}{35(9r_{b}^{2}-9N^{2}+\ell ^{2})}\Big[ %
315r_{b}^{10}+5(7\ell ^{2}-387N^{2})r_{b}^{8}+14N^{2}(369N^{2}-14\ell
^{2})r_{b}^{6}  \notag \\
&&+70N^{4}(7\ell ^{2}-117N^{2})r_{b}^{4}+35N^{6}(333N^{2}-28\ell
^{2})r_{b}^{2}+245N^{8}(9N^{2}-\ell ^{2})\Big]
\end{eqnarray}
(with $r_{b}=r_{b\pm }$). The specific heat is, explicitly,
\begin{eqnarray}
C_{Bolt}(r_{b}=r_{b\pm }) &=&\frac{-256\pi ^{4}\ell
^{10}(326592000N^{6}-3564000\ell ^{2}N^{4}+9450\ell ^{4}N^{2}-7\ell ^{6})}{%
117705877734375N^{8}}  \notag \\
&&\mp \frac{256\pi ^{4}}{117705877734375N^{8}\ell ^{2}\sqrt{\ell
^{4}-900N^{2}\ell ^{2}+8100N^{4}}}\times  \notag \\
&&\Big(34867844010000000000N^{20}-4649045868000000000N^{18}\ell ^{2}  \notag
\\
&&+99007458300000000N^{16}\ell ^{4}+1147912560000000N^{14}\ell ^{6}  \notag
\\
&&+25509168000000N^{12}\ell ^{8}+141717600000\ell
^{10}N^{10}-43040160000N^{8}\ell ^{12}  \notag \\
&&+1318032000\ell ^{14}N^{6}-7136100N^{4}\ell ^{16}+12600N^{2}\ell
^{18}-7\ell ^{20}\Big)
\end{eqnarray}%
The flat space limits for the lower branch solutions are,
\begin{eqnarray*}
I_{Bolt}(r_{b}=r_{b-}) &\rightarrow &\frac{19914752N^{8}\pi ^{4}}{7} \\
S_{Bolt}(r_{b}=r_{b-}) &\rightarrow &19914752N^{8}\pi ^{4} \\
C_{Bolt}(r_{b}=r_{b-}) &\rightarrow &-159318016N^{8}\pi ^{4}
\end{eqnarray*}%
The lower branch Bolt radius goes to $5N$, and so in the flat space limit,
the Bolt mass goes to
\begin{equation}
m_{b}\rightarrow \frac{
5r_{b}^{8}-28r_{b}^{6}N^{2}+70r_{b}^{4}N^{4}-140N^{6}r_{b}^{2}-35N^{8}}{
70r_{b}}=\frac{155584}{35}N^{7}
\end{equation}
As in six dimensions, the upper branch Bolt quantities are all ($\pm $)
infinity in the high temperature limit; the lower branch quantities are all
zero. In ten dimensions, the upper branch solutions for the entropy and
specific heat are thermodynamically stable; the lower branch solutions are
thermodynamically unstable. The plots of these are qualitatively similar to
the six and eight dimensional cases.

\subsection{N\"other conserved quantities}

We take the Taub-bolt solution in 10 dimensions (\ref{10dmetric1}) as the
dynamical metric with mass parameter (\ref{10dmb}) and identification of
period (\ref{10dbetaBolt}). The background is the Taub-NUT metric (\ref%
{10dmetric1}) with mass parameter (\ref{mn10d}). Now the spacetime vector
field is%
\begin{equation}
\xi =\partial _{\tau }+a\partial _{\theta _{1}}+b\partial _{\theta
_{2}}+c\partial _{\theta _{3}}+d\partial _{\theta _{4}}
\end{equation}%
which by definition produces the N\"{o}ther conserved quantity $
Q=m+aJ_{1}+bJ_{2}+cJ_{3}+dJ_{4}$ of the corresponding dynamical metric
relative to the background. Evaluating the superpotential (\ref%
{superpotential}) on the solutions $(g,\bar{g})$ then integrating on a
spatial region $(\tau =\tau _{0},r=r_{0})$ and expanding in the limit $%
r_{0}\rightarrow \infty $, we obtain $Q=Q_{1}+Q_{2}+Q_{3}$, where
\begin{eqnarray*}
Q_{1} &=&{\frac{64\pi^{4}}{\kappa \ell^{2}}}\Bigg( %
2r_{0}^{9}-8N^{2}r_{0}^{7}+\hbox{$4\over
35$}N^{2}(4\ell^{2}+65N^{2})r_{0}^{5}-\hbox{$8\over
35$}N^{4}(12\ell^{2}-85N^{2})r_{0}^{3} \\
&&+\hbox{$2\over
35$}N^{6}(232\ell^{2}-2285N^{2})r_{0}+\hbox{$1\over
5r_b$}\Big(35r_{b}^{10}+5(-45N^{2}+\ell^{2})r_{b}^{8}+14N^{2}(45N^{2}-2
\ell^{2})r_{b}^{6} \\
&&+70N^{4}(-15N^{2}+\ell
^{2})r_{b}^{4}+35N^{6}(45N^{2}-4\ell^{2})r_{b}^{2}+35N^{8}(9N^{2}-\ell^{2}) %
\Big)\Bigg)+O(r_{0}^{-1})
\end{eqnarray*}
\begin{eqnarray*}
Q_{2} &=&{\frac{64\pi^{4}}{35\kappa \ell^{2}r_{b}}}\Big( %
35r_{b}^{10}+5(-45N^{2}+\ell
^{2})r_{b}^{8}+14N^{2}(45N^{2}-2\ell^{2})r_{b}^{6}+70N^{4}(\ell
^{2}-15N^{2})r_{b}^{4}+ \\
&&+35N^{6}(45N^{2}-4\ell
^{2})r_{b}^{2}+128N^{7}(\ell^{2}-10N^{2})r_{b}+35N^{8}(9N^{2}-\ell^{2})\Big) %
+O(r_{0}^{-1})
\end{eqnarray*}
\begin{eqnarray*}
Q_{3} &=&{\frac{64\pi^{4}}{\kappa \ell^{2}}}\Big(-2r_0^{9}+8N^{2}r_0^{7}-
\hbox{$4\over
35$}N^{2}(4\ell^{2}+65N^{2})r_0^{5}+\hbox{$8\over 35$} N^{4}(12
\ell^{2}-85N^{2})r_0^{3}+ \\
&&-\hbox{$2\over
35$}N^{6}(232\ell^{2}-2285N^{2})r_0+\hbox{$128\over
5$}N^{7}(-10N^{2}+\ell^{2})\Big)+O(r_{0}^{-1})
\end{eqnarray*}

We see again that $Q_{1}$ and $Q_{3}$ diverge as $r_{0}\rightarrow \infty $,
while the total conserved quantity $Q$ does not. In fact
\begin{eqnarray*}
Q &=&{\frac{64\pi ^{4}}{\kappa \ell ^{2}}}\Bigg(\hbox{$1024\over
35
$}N^{7}(\ell ^{2}-10N^{2})+ \\
&&\Big(8r_{b}^{9}+\hbox{$8\over 7 $}(-45N^{2}+\ell ^{2})r_{b}^{7}+%
\hbox{$16\over 5$}N^{2}(45N^{2}-2\ell ^{2})r_{b}^{5}+\hbox{$16$}%
N^{4}(-15N^{2}+\ell ^{2})r_{b}^{3} \\
&&+\hbox{$8$}N^{6}(45N^{2}-4\ell ^{2})r_{b}+\hbox{$8\over r_b$}%
N^{8}(9N^{2}-\ell ^{2})\Big)\Bigg)+O(r_{0}^{-1})
\end{eqnarray*}%
As before, no contribution from the angular part of $\xi $ survives, and so $%
J_{1}=J_{2}=J_{3}=J_{4}=0$. Interpreting Q as the relative mass, we find
\begin{equation}
Q=\hbox{$1024\pi^4\over \ka$}(m_{b}-m_{n})
\end{equation}%
As before, from the first law of thermodynamics we have
\begin{equation}
\delta S=20\pi N\delta Q
\end{equation}%
where $S$ is interpreted as the relative entropy of the two solutions. After
a tedious calculation we find

\begin{eqnarray}
S &=&\hbox{$256\pi^5\over 2615686171875\ka
N^7\ell^2$}\Big((-1456542000\ell
^{8}N^{8}+3443737680000000N^{16}-88573500000\ell ^{6}N^{10}  \notag \\
&&-2869781400000\ell ^{4}N^{12}-143489070000000N^{14}\ell ^{2}+3555900\ell
^{12}N^{4}  \notag \\
&&-318573000\ell ^{10}N^{6}-9450\ell ^{14}N^{2}+7\ell
^{16})r_{b}-3443737680000000N^{17}  \notag \\
&&-10975500\ell ^{12}N^{5}-35N\ell ^{16}+16183800\ell
^{10}N^{7}+296544078000000N^{15}\ell ^{2}  \notag \\
&&+984150000\ell ^{8}N^{9}+31886460000\ell ^{6}N^{11}+1594323000000\ell
^{4}N^{13}+39690\ell ^{14}N^{3}\Big)  \notag \\
&=&\hbox{$4\pi\over \kappa$}\Big(S_{Bolt}-S_{NUT}-\frac{216064\pi ^{4}\ell
^{8}}{20503125}\Big)  \label{Snoether10}
\end{eqnarray}

\section{General Solution}

\label{sec:Gensol}

The pattern obtained so far for the Taub-NUT/Bolt-AdS solutions can
straightforwardly be extended to arbitrary even dimensionality \cite{AC}. We
consider in this section the basic thermodynamics of this general solution.

The general form for the Taub-NUT/Bolt-AdS class of metrics for a $U(1)$
fibration over $(S^{2})^{\otimes \left( n-2\right) }$ is
\begin{equation}
ds^{2}=F(r)(d\tau +2N\cos \theta _{i}d\phi _{i})^2+\frac{dr^{2}}{F(r)}%
+(r^{2}-N^{2})(d\theta _{i}^{2}+\sin ^{2}(\theta _{i})d\phi _{i}^{2})
\label{metricd}
\end{equation}
with $i$ summed from $1$ to $n-2$. For convenience we have set $n=d+1$. The
general form for $F(r)$ is \cite{AC}
\begin{equation}
F(r)=\frac{r}{(r^{2}-N^{2})^{k}}\int^{r}\left[ \frac{(s^{2}-N^{2})^{k}}{%
s^{2} }+\frac{2k+1}{\ell ^{2}}\frac{(s^{2}-N^{2})^{k+1}}{s^{2}}\right] ds-%
\frac{2mr }{(r^{2}-N^{2})^{k}}  \label{Frgeneral}
\end{equation}
where $n=2k+2$, and the parameter $m$ is an integration constant. From (\ref%
{metricd}), general expressions for the Ricci scalar and metric determinant
can be found, as well as general expressions for the Ricci scalar and metric
determinant on the boundary (as $r\rightarrow \infty $). From the
expressions for the metric determinant and Ricci scalar of the $d$%
-dimensional boundary metric,
\begin{eqnarray}
\gamma &=&F(r)(r^{2}-N^{2})^{n-2}\prod_{i=1}^{k}\sin ^{2}(\theta _{i})
\label{gammad} \\
R(\gamma ) &=&(n-2)\left[ \frac{1}{(r^{2}-N^{2})}-\frac{F(r)N^{2}}{
(r^{2}-N^{2})^{2}}\right]  \label{Ricscald}
\end{eqnarray}
it is straightforward to show that, by expanding the combinations of these
functions in (\ref{Lagrangianct}) for large $r$, the only finite
contribution from the counter term action (\ref{Lagrangianct}) comes from
the first term as $r\rightarrow \infty $. All of the other terms will either
go to zero or diverge, and hence be used to cancel the divergences from (\ref%
{actbulk}), (\ref{actbound}).

\bigskip The reason for this can be seen as follows. From (\ref{Frgeneral})
we have
\begin{equation*}
F(r)=\sum_{j}r^{2p_{j}}-\frac{2m}{r^{2k+1}}+\text{terms that vanish (even
after integration) as }r\rightarrow \infty \text{ }
\end{equation*}
where the $p_{j}$'s are positive integers. All divergences must therefore be
cancelled by terms that are even powers of $r$\ in the large-$r$\ limit. \
Any term appearing in one of the counterterms that depends on $m$\ must also
depend upon an odd power of $r$, because of the above expansion of $F$. \ It
therefore cannot cancel a divergence. \ Because of dimensionality, all
non-divergent counterterm contributions that depend on $m$\ must be down by
at least one power from $\frac{2m}{r^{2k+1}}$, ie they must at least behave
like $\frac{2m}{r^{2k+2}}$. \ These will all vanish upon integration, since $%
\sqrt{\gamma }\sim r^{2k+1}$in the large-$r$\ limit. Likewise all
non-divergent counterterm contributions that do not depend on $m$\ must be
down by two powers, so they will vanish as well after integrating. We have
explicitly checked these statements up to and including $20$ dimensions.

We thus obtain
\begin{equation}
I_{ct\text{finite}}=-\frac{(n-2)(4\pi )^{(n-2)/2}\beta }{8\pi }m
\label{ictfinite}
\end{equation}
for the general finite contribution from the counter-term action.

For the finite contribution from the boundary action (\ref{actbound}), a
general expression for the trace of the extrinsic curvature $\Theta $ can be
obtained from the metric (\ref{metricd})%
\begin{equation}
\Theta =\frac{F^{\prime }(r)}{2\sqrt{F(r)}}+(n-2)\frac{r\sqrt{F(r)}}{
(r^{2}-N^{2})}  \label{TrKexgen}
\end{equation}
Expanding $\sqrt{\gamma }\Theta $ for large $r$, the general finite
contribution from the boundary action is
\begin{equation}
I_{\partial B\text{finite}}=\frac{(n-1)(4\pi )^{(n-2)/2}\beta }{8\pi }m
\label{Ibfinite}
\end{equation}

Finally, using the general expressions for the metric determinant and $R(g)$%
,
\begin{eqnarray}
g &=&(r^{2}-N^{2})^{(n-2)}\prod_{i=1}^{k}\sin ^{2}(\theta _{i}) \\
R(g) &=&-\frac{n(n-1)}{\ell ^{2}}
\end{eqnarray}
the general expression for the finite $I_{B}$ term can be found. (Note that $%
\prod \sin ^{2}(\theta _{i})$ from both $g$ and $\gamma $ contribute to the
volume element $(4\pi )^{k}$). Substituting these into (\ref{actbulk}), we
get
\begin{equation}
I_{B}=\frac{(n-1)(4\pi )^{(n-2)/2}\beta }{8\pi \ell ^{2}}\int
dr(r^{2}-N^{2})^{(n-2)/2}
\end{equation}
The integrand here can be expanded using the binomial theorem and
integrated, term by term, for the range $r=(r_{+},r^{\prime })$, where $%
r^{\prime }\rightarrow r\rightarrow \infty $, and so the general expression
for the finite contribution from the bulk action is
\begin{equation}
I_{B\text{finite}}=-\frac{(n-1)(4\pi )^{(n-2)/2}\beta }{8\pi \ell
^{2}} \sum_{i=0}^{(n-2)/2}\left( {\frac{(n-2)}{2} \atop i }
\right) (-1)^{i}N^{2i}\left[ \frac{r_{+}^{n-2i-1}}{n-2i-1}\right]
\label{Ibulkfinite}
\end{equation}
Adding together (\ref{Ibulkfinite}), (\ref{Ibfinite}) and (\ref{ictfinite})
as in (\ref{action}), the general expression for the action in $n=d+1$
dimensions is
\begin{equation}
I=\frac{(4\pi )^{(n-2)/2}\beta }{8\pi \ell ^{2}}\left[ m\ell
^{2}-(n-1)\sum_{i=0}^{(n-2)/2}\left( {\frac{(n-2)}{2} \atop i }
\right) (-1)^{i}N^{2i}\frac{r_{+}^{(n-2i-1)}}{(n-2i-1)}\right]
\label{actiongeneral}
\end{equation}

The expression (\ref{Mcons}) can also be used to find the conserved mass for
the Taub-NUT/Bolt-AdS class of metrics in any dimension. The divergence
cancellations take place in a manner analogous to that described above for
the action. We find
\begin{equation}
\mathfrak{M}=\frac{(n-2)(4\pi )^{(n-2)/2}m}{8\pi }  \label{MassGen}
\end{equation}
and we have explicitly checked that it satisfies the first law of
thermodynamics for $n=4\ldots 20$ (even dimensions) for both the NUT and
Bolt solutions (see below).

\subsection{General NUT Solution}

Using the binomial expansion on (\ref{Frgeneral}), and the condition $%
F(r=N)=0$, a general formula
\begin{subequations}
\begin{eqnarray}
m_{n} &=&\frac{N^{n-3}}{2\ell ^{2}}\Big[\ell ^{2}-nN^{2}\Big] %
\sum_{i=0}^{(n-2)/2}\left( {\frac{(n-2)}{2} \atop i } \right)
\frac{
(-1)^{i}}{n-2i-3}  \label{mngeneral1} \\
&=&\frac{N^{n-3}}{\sqrt{\pi }\ell ^{2}}\Big[\ell ^{2}-nN^{2}\Big]\frac{
\Gamma \left( \frac{5-n}{2}\right) \Gamma \left( \frac{n}{2}\right) }{\left(
n-3\right) }  \label{mngeneral2}
\end{eqnarray}
for the NUT mass can be obtained. From this, and from the identity
\end{subequations}
\begin{equation}
\sum_{i=0}^{k+1}\left( { k+1 \atop i } \right)
\frac{(-1)^{i}}{(2k-2i+1)} =-\left( \frac{2k+2}{2k+1}\right)
\sum_{i=0}^{k}\left( { k \atop i } \right)
\frac{(-1)^{i}}{(2k-2i-1)}
\end{equation}
which can be proven by induction, we see that the general form of the NUT
mass will remain the same ($\propto \left[ \ell ^{2}-nN^{2}\right] $), since
the sum in (\ref{mngeneral1}) is a constant. However the overall sign will
change (in the large-$\ell /N$ limits) alternating between the (even)
dimensionalities.

Using (\ref{actiongeneral}), the general form for the NUT action is
\begin{equation}
I_{NUT}=\frac{(4\pi )^{(n-2)/2}\beta }{16\pi ^{3/2}\ell ^{2}}\Gamma \left(
\frac{3-n}{2}\right) \Gamma \left( \frac{n}{2}\right) N^{n-3}\left[
(n-2)N^{2}-\ell ^{2}\right]  \label{INUTgeneral}
\end{equation}
where, as with $m_{n}$, there is an alternating sign between even
dimensionalities, though its overall form remains the same.

It is now an easy matter to find the general form of the NUT entropy, using (%
\ref{MassGen}) and (\ref{entropy}). We find that
\begin{equation}
S_{NUT}=\frac{(4\pi )^{(n-2)/2}\beta }{16\pi ^{3/2}\ell ^{2}}\Gamma \left(
\frac{3-n}{2}\right) \Gamma \left( \frac{n}{2}\right) N^{n-3}\left[
(n-1)(n-2)N^{2}-(n-3)\ell ^{2}\right]  \label{SNUTgeneral}
\end{equation}
Likewise a general expression for the specific heat can be found, using the
relation $C=-\beta \partial _{\beta }S$ and the general form
\begin{equation}
\beta =2n\pi N
\end{equation}
for $\beta $; we obtain
\begin{equation}
C_{NUT}=-\frac{(4\pi )^{(n-2)/2}\beta }{16\pi ^{3/2}\ell ^{2}}\Gamma \left(
\frac{3-n}{2}\right) \Gamma \left( \frac{n}{2}\right) N^{n-3}\left[
n(n-1)(n-2)N^{2}-(n-2)(n-3)\ell ^{2}\right]  \label{CNUTgeneral}
\end{equation}

Examining (\ref{INUTgeneral}), (\ref{SNUTgeneral}), and (\ref{CNUTgeneral}),
we see that the NUT quantities all vanish in the high temperature limit as $%
N\rightarrow 0$. Notice also that the $\Gamma \left( {\textstyle\frac{3-n}{2}
}\right) $ will produce a minus sign for every odd value of $k$ (recall $
n=2k+2$). This will mean that, for odd $k$ ($n=4,8,12,\ldots $), the entropy
and specific heat will both be positive in the following range:
\begin{equation}
\ell \sqrt{\frac{(n-3)}{n(n-1)}}<N<\ell \sqrt{\frac{(n-3)}{(n-1)(n-2)}}
\end{equation}
Hence, for odd $k$, the NUT solutions will be thermally stable in this
range, which becomes increasingly narrow with increasing dimensionality. For
even $k$ ($n=6,10,14,\ldots $), however, no minus sign is produced, meaning
that the entropy will be positive for $N>\ell \sqrt{{\textstyle\frac{n-3}{
(n-1)(n-2)}}}$, and the specific heat will be positive for $N<\ell \sqrt{{\ %
\textstyle\frac{(n-3)}{n(n-1)}}}$. Since the second value will always be
larger than the first, this means there is no range in which both will be
positive, and thus for all even $k$, the NUT solutions will be thermally
unstable.

Also, the general flat space action, entropy and specific heat will be given
by
\begin{eqnarray}
I_{NUT} & \rightarrow & - \frac{(4\pi )^{(n-2)/2} \beta }{16 \pi^{3/2} }
\Gamma \left( \frac{3-n}{2} \right) \Gamma \left( \frac{n}{2} \right) N^{n-3}
\\
S_{NUT} & \rightarrow & - \frac{ (4 \pi )^{(n-2)/2} \beta}{16 \pi^{3/2} }
\Gamma \left( \frac{3-n}{2} \right) \Gamma \left( \frac{n}{2} \right)
N^{n-3} (n-3) \\
C_{NUT} & \rightarrow & \frac{ (4 \pi )^{(n-2)/2} \beta }{16 \pi^{3/2} }
\Gamma \left( \frac{3-n}{2} \right) \Gamma \left( \frac{n}{2} \right)
N^{n-3} (n-2)(n-3)
\end{eqnarray}
Notice that the $\Gamma \left( {\textstyle\frac{3-n}{2}} \right) \Gamma
\left( {\textstyle\frac{n}{2}} \right)$ will give the same sign change with
an increase in (even) dimension in the flat space limit values for $%
S_{NUT},~C_{NUT}$. This means that in any dimension, the asymptotically
locally flat pure NUT solutions will always be thermally unstable, for any
dimension.

\subsection{General Bolt Solution}

The conditions for arbitrary ($n=d+1$) dimensions that give a regular Bolt
at $r=r_{b}>N$ are $F(r_{b})=0$ and $F^{\prime }(r_{b})={\frac{2}{nN}}$. \

The first condition implies from (\ref{Frgeneral}) that
\begin{equation}
m_{b}=\frac{1}{2}\left[ \sum_{i=0}^{k}\left( {k \atop i } \right)
\frac{
(-1)^{i}N^{2i}r_{b}^{2k-2i-1}}{(2k-2i-1)}+\frac{(2k+1)}{\ell ^{2}}
\sum_{i=0}^{k+1}\left( {k+1 \atop i }\right) \frac{
(-1)^{i}N^{2i}r_{b}^{2k-2i+1}}{(2k-2i+1)}\right] \label{mbgeneral}
\end{equation}
from which a general expression for the Bolt action can be obtained
\begin{eqnarray}
I_{Bolt} &=&\frac{(4\pi )^{(n-2)/2}\beta }{16\pi \ell ^{2}}\Bigg\{\frac{
(2k+1)(-1)^{k}N^{2k+2}}{r_{b}}  \label{IBoltgeneral} \\
&&+\sum_{i=0}^{k}\left( {k \atop i }\right) (-1)^{i}N^{2i}r_{b}^{2k-2i} %
\left[ \frac{\ell ^{2}}{r_{b}(2k-2i-1)}-\frac{r_{b}(2k+1)(k-2i+1)}{
(2k-2i+1)(k-i+1)}\right] \Bigg\}  \notag
\end{eqnarray}%
by substituting $m=m_{b}$ in (\ref{actiongeneral}). Next, using eq. (\ref%
{entropy}), and substituting $m=m_{b}$ into (\ref{MassGen}), we find
\begin{eqnarray}
S_{Bolt} &=&\frac{(4\pi )^{(n-2)/2}\beta }{16\pi \ell ^{2}}\Bigg\{\frac{
(2k-1)(2k+1)(-1)^{k}N^{2k+2}}{r_{b}}  \label{SGeneralBolt} \\
&&+\sum_{i=0}^{k}\left( {k \atop i }\right) (-1)^{i}N^{2i}r_{b}^{2k-2i} %
\left[ \frac{(2k-1)\ell ^{2}}{r_{b}(2k-2i-1)}+\frac{
(2k+1)(2k^{2}+3k-2i+1)r_{b}}{(2k-2i+1)(k-i+1)}\right] \Bigg\}  \notag
\end{eqnarray}
for the general expression for the Bolt entropy in $n=d+1$ dimensions. The
explicit expression for the specific heat is extremely cumbersome; we shall
not include it here.

\begin{figure}[tbp]
\begin{minipage}[c]{.45\textwidth}
         \centering
         \includegraphics[width=\textwidth]{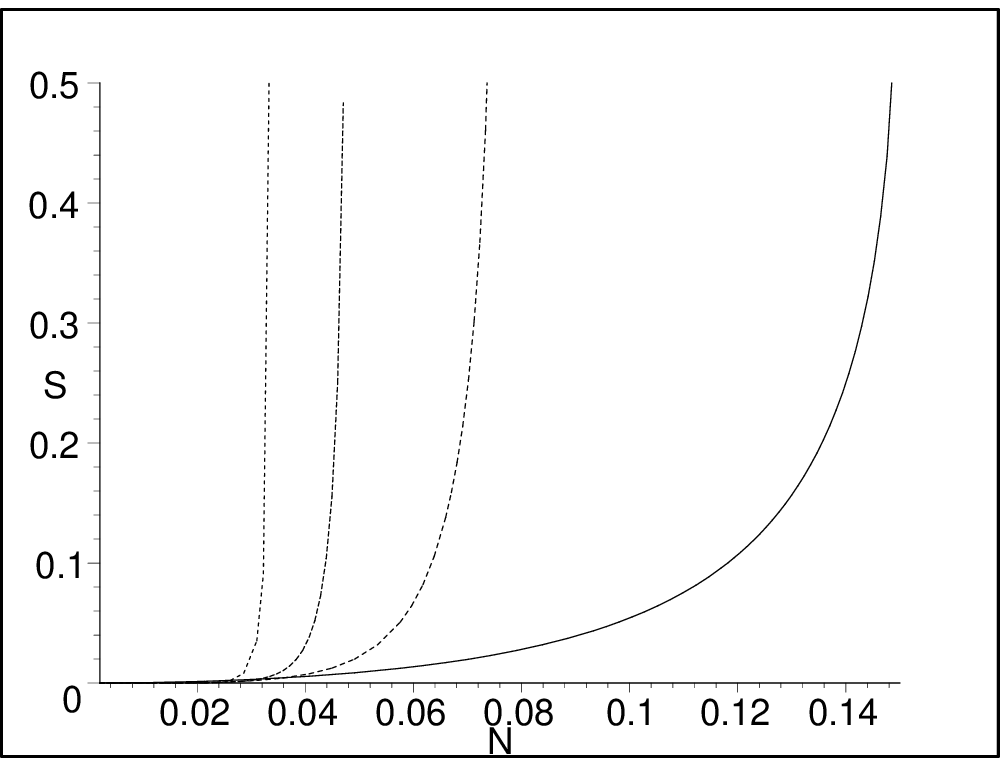}
         \caption{Plot of the Relative entropies for (from right to
left) 4 to 10 dimensions}
         \label{RelativeSPlot}
     \end{minipage}
\begin{minipage}[c]{.40\textwidth}
\end{minipage}
\begin{minipage}[c]{.45\textwidth}
         \centering
         \includegraphics[width=\textwidth]{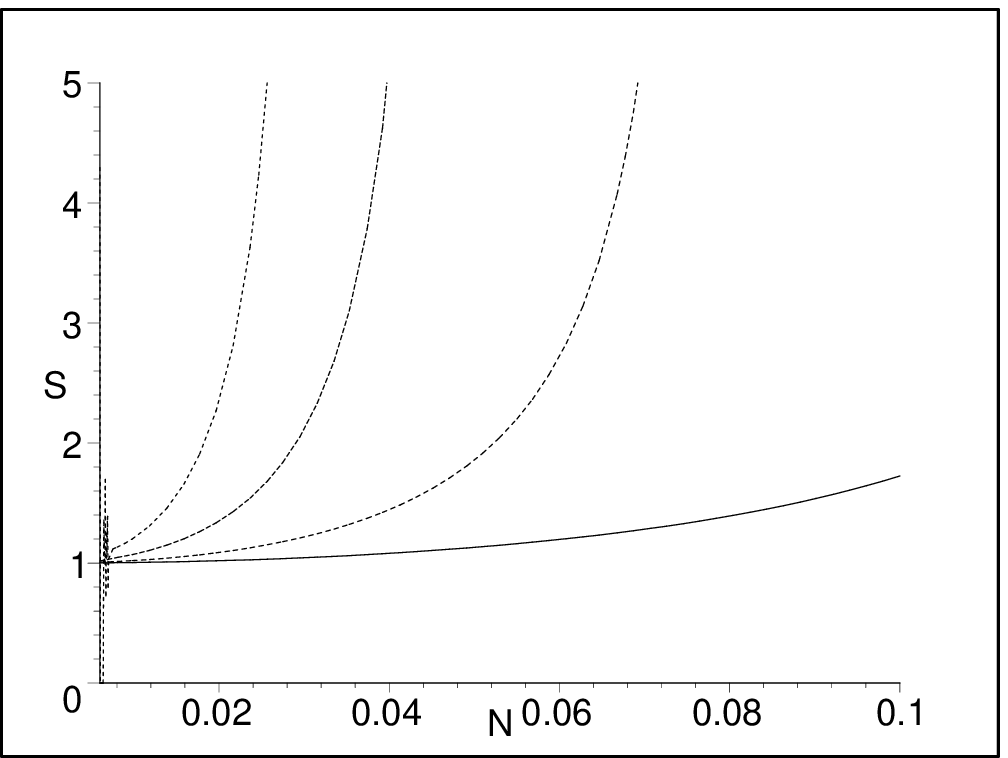}
         \caption{Plot of the Re-scaled Relative entropies for (from
right to left) 4 to 10 dimensions}
         \label{RescaledRelSPlot}
     \end{minipage}
\end{figure}

An analysis of the general bolt case is somewhat awkward, though it is
possible to deduce some general trends. For example, we see that the
relative entropies ($S_{Rel}=S_{Bolt}(r=r_{b-})-S_{NUT}$) increase (with
increasing $N$) faster as we increase the number of dimensions. This can be
seen in Figure \ref{RelativeSPlot}, where we plot the relative entropies
from four to ten dimensions. \ From small values of $N$, the entropy
increases with decreasing dimensionality. However this rapidly changes once $%
N$\ becomes sufficiently large, in which case the entropy rapidly increases
with increasing dimensionality like $N^{d-2}$. \ \ In Figure \ref%
{RescaledRelSPlot}, we have plotted the relative entropies with the
prefactor of $N^{d-2}$\ scaled out (so that all relative entropies are unity
at $N=0$) -- we see that entropy still increases with increasing
dimensionality.

\section{Conclusions}

Our general investigation of the thermodynamics of the $\left( d+1\right) $
-dimensional class of Taub-NUT-AdS metrics has yielded a number of specific
results which we recapitulate here. We have shown consistency between the
Noether-charge and counterterm approaches despite the a-priori distinction
between the two methods. \ The surprising results \cite{MannMisner,EJM}
obtained in four dimensions carry over to higher dimensionalities, with
certain qualifications. \ For $4k$ dimensions there is a finite range of $%
N/\ell $ for which both the entropy and specific heat are positive
\begin{equation}
\ell \sqrt{\frac{(n-3)}{n(n-1)}}<N<\ell \sqrt{\frac{(n-3)}{(n-1)(n-2)}}
\end{equation}
for the NUT cases. Outside of this range at least one of these quantities
becomes negative. \ In $4k+2$ dimensions no such range exists: either the
entropy or specific heat (or both) is negative for a given value of $N/\ell $
. \ \ In contrast to this, for all bolt cases there exists a
thermodynamically stable region in which both the entropy and specific heat
are positive, as well as a thermodynamically unstable region. The general
behaviour is illustrated in figs. \ref{6dSCBoltplot}, \ref{6dSBmPlot}, \ref%
{6dCBmPlot}. \ For large $d$, the magnitudes of all thermodynamic quantities
diverge, with $C_{\text{NUT}}\sim dS_{\text{NUT}}$. $\sim d^{2}I_{\text{NUT}
} $, with an overall factor of $N^{d-1}$ common to each. \ \

It is difficult to know how to interpret such results. \ Since they depend
sensitively on the analytic continuation of the NUT parameter to imaginary
values, it is not clear to what extent they can be carried over to a
Lorentzian framework. \ The appearance of negative entropy for a rather
broad range of values of $N/\ell $ in the NUT cases is particularly
troubling. We have taken this to mean that such spacetimes are
thermodynamically unstable. However we find from both the Noether charge and
counterterm approaches that the entropy difference between the NUT and bolt
cases is always positive. \ It would be useful to confirm (or refute) these
results by other methods that do not depend upon analytic continuation.

We close with some comments on the relationship between the Noether-charge
and counterterm approaches. \ The former approach invokes a background
spacetime in order to make all quantities well-defined. \ A Noether-charge
result for $(d+1)$\ dimensions is rather tedious to obtain. However based on
the earlier results for $4,6,8$\ and $10$\ dimensions, we conjecture that
\begin{equation*}
Q=\frac{(d-1)(4\pi )^{(d-1)/2}}{8\pi }\left( m_{b}-m_{n}\right) =\mathfrak{M}%
_{bolt}-\mathfrak{M}_{NUT}
\end{equation*}%
The entropy is then computed up to an overall constant of integration by
assuming the first law to be valid, in which case we then obtain
\begin{equation*}
S_{Q}=\hbox{$4\pi\over \kappa$}(S_{Bolt}-S_{NUT})+c_{l}
\end{equation*}%
where $c_{l}$\ is an $N$-independent constant. Although it can have no
effect on the first law of thermodynamics, (since this law is only sensitive
to changes of entropy) it is important in ensuring that the ground state of
the theory (ie the small $N$) limit is not degenerate. \

\begin{figure}[tbp]
\begin{minipage}[c]{.45\textwidth}
         \centering
         \includegraphics[width=\textwidth]{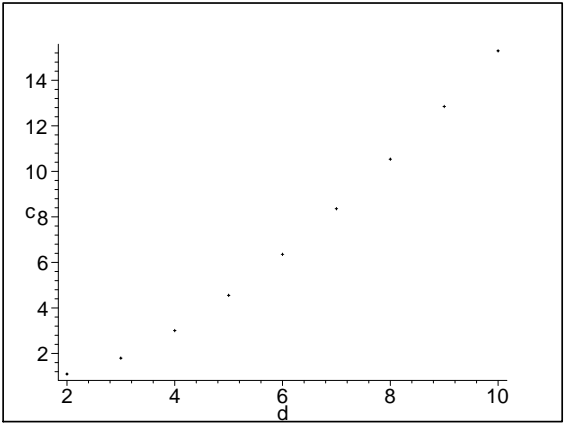}
         \caption{Plot of $-\ln(C_{d+1})$ vs. $d$
from 4 to 20 dimensions}
         \label{Lorconstplot}
     \end{minipage}
\end{figure}

We can see this as follows. From eqs. (\ref{MassGen}), (\ref{mngeneral2})
and (\ref{SNUTgeneral}) that if we define a quantity $\hat{S}$\ such that%
\begin{equation}
d\hat{S}=\beta d\mathfrak{M}_{NUT}=\beta \frac{d\mathfrak{M}_{NUT}}{dN}dN
\label{Shatnut}
\end{equation}%
then upon inserting (\ref{mngeneral2}) into this relation and integrating we
find%
\begin{equation}
\hat{S}=\frac{(4\pi )^{(n-2)/2}\beta }{16\pi ^{3/2}\ell ^{2}}\Gamma \left(
\frac{3-n}{2}\right) \Gamma \left( \frac{n}{2}\right) N^{n-3}\left[
(n-1)(n-2)N^{2}-(n-3)\ell ^{2}\right] =S_{NUT}  \label{Shatnut2}
\end{equation}%
which is the general formula for the pure NUT entropy. \ Consequently it is
the Bolt solutions that have the additional constant, since the relation
\begin{equation}
dS_{Q}=\beta dQ=\beta d\mathfrak{M}_{bolt}-\beta d\mathfrak{M}_{NUT}
\label{firstlawrel}
\end{equation}%
implies that
\begin{equation}
S_{bolt}+c_{l}=\int \beta \frac{d\mathfrak{M}_{bolt}}{dN}dN  \label{Sbolt}
\end{equation}%
It is easy to obtain this constant from the Gibbs-Duhem relation (\ref%
{entropy}) by inserting the explicit functional form of $r_{b}(N)$\ into
this relation and isolating the $N$-independent term. This has the form%
\begin{equation}
c_{l}=\frac{\pi }{\ell ^{2}}C_{d+1}\left( -\pi \ell ^{2}\right) ^{\left(
d+1\right) /2}  \label{cl}
\end{equation}%
in $\left( d+1\right) $\ dimensions, where we have computed \ these
coefficients up to $d=19$:%
\begin{eqnarray}
C_{4} &=&1/3,\text{ \ \ \ }C_{6}=\frac{112}{675},\text{ \ \ \ \ }C_{8}=\frac{%
17}{343},\text{ \ \ \ \ }C_{10}=\frac{216064}{20503125}\text{ \ \ \ \ \ \ \ }%
C_{12}=\frac{22672}{13045131}\text{ \ \ \ \ }  \notag \\
C_{14} &=&\frac{930414592}{3975084764287},\text{ \ \ \ }C_{16}=\frac{58253}{%
2187000000},\text{ \ \ \ \ }C_{18}=\frac{7092488372224}{2702551358937608649},%
\text{ \ \ \ }  \label{Cds} \\
\text{\ }C_{20} &=&\frac{28736819456}{126049881944921875}  \notag
\end{eqnarray}%
\ and these constants are the same for both branches $r_{b+}$\ and \ $r_{b-}$%
\ . \ We have plotted them in fig. (\ref{Lorconstplot}).

It is straightforward to check that the constant in eq. (\ref{cl}) ensures
that the Bolt entropy vanishes in the small-$N$\ limit. \ This ensures that
there is no ground-state ($N=0$) degeneracy in the associated CFT. \ The
Noether-charge method, of course, is insensitive to \ any such $N$%
-independent constant, as is clear from eq. (\ref{Sbolt}). \ An off-shell
comparison between the two approaches should shed further light on these and
other issues.

\bigskip

This work was supported in part by the Natural Sciences and Engineering
Research Council of Canada and by INdAM-GNFM.

\begin{table}[!hbp]
\caption{Summary of NUT quantities }
\label{summarytableNUT}\centering
\begin{tabular}{|c|c|c|c|c|c|}
\hline
Dim. & Period. & Mass & Action & Entropy & Specific Heat \\ \hline
&  &  &  &  &  \\
4 & $8\pi N$ & $\frac{N(\ell^{2}-4N^{2})}{\ell^{2}} $ & $\frac{ 4\pi N^{2}
(\ell^{2}-2 N^{2})}{ \ell^{2}} $ & $\frac{4\pi N^{2} (\ell^{2} -6 N^{2})}{%
\ell^{2}} $ & $\frac{8\pi N^{2} (12N^{2} -\ell^{2})}{\ell^{2} } $ \\
&  &  &  &  &  \\ \hline
&  &  &  &  &  \\
6 & $12 \pi N$ & $\frac{4N^{3}(6N^{2}-\ell^{2})}{3\ell^{2}} $ & $\frac{%
32\pi^{2}N^{4}(4N^{2}-\ell^{2})}{\ell^{2}} $ & $\frac{32%
\pi^{2}N^{4}(20N^{2}-3\ell^{2})}{\ell^{2}} $ & $\frac{384\pi^{2}N^{4}(%
\ell^{2}-10N^{2})}{\ell^{2}} $ \\
&  &  &  &  &  \\ \hline
&  &  &  &  &  \\
8 & $16 \pi N $ & $\frac{8N^{5}(\ell^{2}-8N^{2})}{5\ell^{2}} $ & $\frac{
1024\pi^{3}N^{6}(\ell^{2}-6N^{2})}{5\ell^{2}} $ & $\frac{1024\pi^{3}N^{6}(5
\ell^{2}-42N^{2})}{5\ell^{2}} $ & $\frac{6144\pi^{3}N^{6}(56N^{2}-5\ell^{2})%
}{ 5\ell^{2}} $ \\
&  &  &  &  &  \\ \hline
&  &  &  &  &  \\
10 & $20 \pi N $ & $\frac{ 64 N^{7}(10 N^{2} - \ell^{2})}{35\ell^{2} } $ & $%
\frac{ 8192 \pi^{4} N^{8} (8N^{2} - \ell^{2})}{7 \ell^{2} } $ & $\frac{ 8192
\pi^{4} N^{8} (72 N^{2}-7\ell^{2})}{7 \ell^{2} } $ & $\frac{ 65536 \pi^{4}
N^{8} (7 \ell^{2}-90N^{2})}{ 7\ell^{2}} $ \\
&  &  &  &  &  \\ \hline
&  &  &  &  &  \\
n & $2n\pi N$ & $\frac{A [\ell^{2}-nN^{2}]}{\sqrt{\pi }\ell^{2} \left(
n-3\right) } $ & $\frac{B [ (n-2)N^{2} - \ell^{2} ] }{16 \pi^{3/2} \ell^{2} }
$ & $\frac{ B [ (n-1)(n-2)N^{2} - (n-3)\ell^{2} ] }{16\pi^{3/2} \ell^{2}} $
& $\frac{ B [(n-2)(n-3)\ell^{2} - n(n-1)(n-2) N^{2} ]}{16\pi^{3/2} \ell^{2}}
$ \\
&  &  &  &  &  \\ \hline
& \multicolumn{4}{l}{} &  \\
& \multicolumn{4}{l}{$A=N^{n-3} \Gamma \left( \frac{5-n}{2} \right) \Gamma
\left( \frac{n}{2} \right) $, ~~~$B = (4\pi)^{(n-2)/2} \beta \Gamma \left(
\frac{3-n}{2} \right) \Gamma \left( \frac{n}{2} \right) N^{n-3} $} &  \\
& \multicolumn{4}{l}{} &  \\ \hline
\end{tabular}%
\end{table}

\begin{table}[!hbp]
\caption{Summary of Bolt quantities }
\label{summarytableBolt}\centering
\begin{tabular}{|c|c|c|c|}
\hline
Dim. & Mass & Action & Entropy \\ \hline
&  &  &  \\
4 & $\frac{ r_b^{4} + ( \ell^{2} - 6 N^{2} ) r_b^{2} + N^{2} ( \ell^{2} - 3
N^{2})}{ 2\ell^{2} r_b} $ & $\frac{ - \pi (r_b^{4} - \ell^{2} r_b^{2} +
N^{2}( 3N^{2} - \ell^{2}) ) }{3r_b^{2} - 3N^{2} + \ell^{2} } $ & $\frac{ \pi
(3r_b^{4} + ( \ell^{2} - 12 N^{2} ) r_b^{2} + N^{2} (\ell^{2} - 3N^{2}) ) }{%
3r_b^{2} - 3N^{2} + \ell^{2}} $ \\
&  &  &  \\ \hline
&  &  &  \\
6 & $%
\begin{array}{c}
\frac{1}{6r_b \ell^{2} } \Big[ 3r_b^{6} + (\ell^{2} - 15N^{2})r_b^{4} \\
- 3N^{2}(2\ell^{2} - 15N^{2})r_{b}^{2} \\
- 3N^{4}(\ell^{2}-5N^{2}) \Big]%
\end{array}
$ & $%
\begin{array}{c}
\frac{-4\pi^{2}}{3(5r_{b}^{2}-5N^{2}+\ell^{2})} \Big[ 3r_{b}^{6} \\
- (5N^{2} + \ell^{2})r_{b}^{4} \\
- 3N^{2}(5N^{2} - 2\ell^{2})r_{b}^{2} \\
+ 3N^{4}(\ell^{2}-5N^{2}) \Big]%
\end{array}
$ & $%
\begin{array}{c}
\frac{4\pi^{2}}{3(5r_{b}^{2} - 5N^{2} + \ell^{2})} \Big[15r_{b}^{6} \\
- (65N^{2}-3\ell^{2})r_{b}^{4} \\
+ 3N^{2}(55N^{2} - 6\ell^{2})r_{b}^{2} \\
+ 9N^{4}(5N^{2} - \ell^{2}) \Big]%
\end{array}
$ \\
&  &  &  \\ \hline
&  &  &  \\
8 & $%
\begin{array}{c}
\frac{1}{10\ell^{2}r_{b} } \Big[ 5r_{b}^{8} \\
+ (\ell^{2} - 28N^{2})r_{b}^{6} \\
+ 5N^{2}(14N^{2} - \ell^{2})r_{b}^{4} \\
+ 5N^{4}(3\ell^{2} - 28N^{2})r_{b}^{2} \\
+ 5N^{6}(\ell^{2} - 7N^{2}) \Big]%
\end{array}
$ & $%
\begin{array}{c}
\frac{-16\pi^{3}}{5(7r_{b}^{2} - 7N^{2} + \ell^{2} )} \Big[ 5r_{b}^{8} \\
- (\ell^{2} + 14N^{2})r_{b}^{6} \\
+ 5N^{2}r_{b}^{4}\ell^{2} \\
- 5N^{4}(3\ell^{2} - 14N^{2})r_{b}^{2} \\
- 5N^{6}(\ell^{2} - 7N^{2}) \Big]%
\end{array}
$ & $%
\begin{array}{c}
\frac{16\pi^{3}}{5(7r_{b}^{2} - 7N^{2} + \ell^{2})} \Big[ 35r_{b}^{8} \\
+ (5\ell^{2} - 182N^{2})r_{b}^{6} \\
- 5N^{2}(5\ell^{2} - 84N^{2})r_{b}^{4} \\
- 5N^{4}(154N^{2} - 15\ell^{2})r_{b}^{2} \\
+ 25N^{6}(\ell^{2} - 7N^{2}) \Big]%
\end{array}
$ \\
&  &  &  \\ \hline
&  &  &  \\
10 & $%
\begin{array}{c}
\frac{1}{70\ell^{2} r_b} \Big[35 r_b^{10} \\
+ 5(\ell^{2} - 45 N^{2} ) r_b^{8} \\
+ 14 N^{2} (45 N^{2} - 2 \ell^{2} )r_b^{6} \\
+ 70N^{4} (\ell^{2} - 15 N^{2})r_b^{4} \\
+ 35 N^{6} (45 N^{2} - 4 \ell^{2} )r_b^{2} \\
+ 35N^{8} (9 N^{2} - \ell^{2} ) \Big]%
\end{array}
$ & $%
\begin{array}{c}
\frac{ -64 \pi^{4} }{35(9 r_b^{2} - 9 N^{2} + \ell^{2} )} \Big[ 35 r_b^{10}
\\
- 5(\ell^{2} + 27 N^{2})r_b^{8} \\
+ 14 N^{2} (2 \ell^{2} + 9 N^{2})r_b^{6} \\
- 70 N^{4} (\ell^{2} - 3 N^{2})r_b^{4} \\
- 35 N^{6} (27 N^{2} - 4 \ell^{2} ) r_b^{2} \\
- 35 N^{8} (9 N^{2} - \ell^{2}) \Big]%
\end{array}
$ & $%
\begin{array}{c}
\frac{ 64 \pi^{4} }{ 35( 9r_b^{2} - 9 N^{2} + \ell^{2} ) } \Big[ 315 r_b^{10}
\\
+ 5( 7 \ell^{2} - 387 N^{2} )r_b^{8} \\
+ 14 N^{2} (369 N^{2} - 14 \ell^{2} )r_b^{6} \\
+ 70 N^{4} ( 7 \ell^{2} - 117 N^{2} )r_b^{4} \\
+ 35 N^{6} (333 N^{2} - 28 \ell^{2} )r_b^{2} \\
+ 245 N^{8}( 9 N^{2} - \ell^{2}) \Big]%
\end{array}
$ \\
&  &  &  \\ \hline
&  &  &  \\
n & $%
\begin{array}{c}
\frac{1}{2} \Bigg[ \sum_1 A_1 \frac{(-1)^i N^{2i} r_b^{2k-2i-1}}{(2k-2i-1)}
\\
+ \frac{(2k+1)}{\ell^{2} } \sum_2 A_2 \frac{(-1)^i N^{2i} r_b^{2k-2i+1}}{
(2k-2i+1)} \Bigg]%
\end{array}
$ & $%
\begin{array}{c}
V \Bigg\{ \frac{ (2 k + 1) (-1)^k N^{2 k + 2} }{r_b } \\
+ \sum_3 A_3 \Big[ \frac{ \ell^{2} }{r_b (2 k - 2i - 1)} \\
- \frac{r_b (2k + 1) (k - 2i + 1)}{(2k - 2i + 1) (k - i + 1) } \Big] \Bigg\}%
\end{array}
$ & $%
\begin{array}{c}
V \Bigg\{ \frac{ (2k - 1) (2k + 1) (-1)^k N^{2k+2}}{r_b} \\
+ \sum_3 A_3 \Big[ \frac{ (2k-1)\ell^{2} }{ r_b (2k-2i-1)} \\
+ \frac{ (2k+1) (2k^{2} + 3k - 2i + 1)r_b}{(2k - 2i + 1)(k-i+1)} \Big] %
\Bigg\}%
\end{array}
$ \\
&  &  &  \\ \hline
& \multicolumn{2}{l}{~~} &  \\
& \multicolumn{3}{l}{$\sum_1 A_1 = \sum_{i=0}^k \left( { k \atop i
} \right) $, ~~$\sum_2 A_2 = \sum_{i=0}^{k+1} \left( { k+1 \atop i
} \right) $, ~~$\sum_3 A_3 = \sum_{i=0}^k \left( { k \atop i }
\right)
(-1)^i N^{2i} r_b^{2k-2i} $,} \\
& \multicolumn{2}{l}{~~} &  \\
& \multicolumn{2}{l}{$V = \frac{ (4\pi)^{(n-2)/2} \beta }{ 16\pi \ell^{2} } $%
} &  \\
& \multicolumn{2}{l}{~~} &  \\ \hline
\end{tabular}%
\end{table}

\end{document}